\documentclass[12pt,titlepage]{article}
\usepackage{amsmath}
\usepackage{amssymb}
\usepackage{graphicx}
\usepackage{caption2}
\usepackage{amsfonts}
\usepackage{cite}

\newcommand{\be}{\begin{equation}}
\newcommand{\ee}{\end{equation}}
\newcommand{\bea}{\begin{eqnarray}}
\newcommand{\eea}{\end{eqnarray}}
\newcommand{\bef}{\begin{figure}}
\newcommand{\ef}{\end{figure}}
\newcommand{\bt}{\begin{tabular}}
\newcommand{\et}{\end{tabular}}
\newcommand{\bno}{\begin{enumerate}}
\newcommand{\eno}{\end{enumerate}}

\def\3{\ss}

\catcode`\"=\active
\def"{\accent'177}

\begin{document}

\begin{center}

\pagestyle{myheadings}

{\bf \large On the reliability of computed chaotic solutions \\
of nonlinear differential equations}

Shijun Liao \footnote{Correspondence.  Email: sjliao@sjtu.edu.cn }

{ State Key Laboratory of Ocean Engineering\\
School of Naval Architecture, Ocean and Civil Engineering\\
Shanghai Jiao Tong University, Shanghai 200030, China}

\end{center}

 {{\bf Abstract} \em In this paper a new concept, namely the critical predictable time
 $T_c$,
is introduced to give a more precise description of computed
chaotic solutions of nonlinear differential equations:  it is
suggested that computed chaotic solutions are unreliable and
doubtable when $ t > T_c$.  This provides us a strategy to detect
reliable solution from a given computed result. In this way, the
computational phenomena, such as computational chaos (CC),
computational periodicity (CP) and computational prediction
uncertainty, which are mainly based on long-term properties of
computed time series, can be completely avoided. Using this
concept, the famous conclusion ``accurate long-term prediction of
chaos is impossible'' should be replaced by a more precise
conclusion that ``accurate prediction of chaos beyond the critical
predictable time $T_c$ is impossible''.  So, this concept also
provides us a time-scale to determine whether or not a particular
time is long enough for a given nonlinear dynamic system. Besides,
the influence of data inaccuracy and  various numerical schemes on
the critical predictable time is investigated in details by using
symbolic computation software as a tool.  A reliable chaotic
solution of Lorenz equation in a rather large interval $0 \leq t <
1200$ non-dimensional Lorenz time units is obtained for the first
time.  It is found that the precision of initial condition and
computed data at each time-step, which is mathematically necessary
to get such a reliable chaotic solution in such a long time, is so
high that it is physically impossible due to the Heisenberg
uncertainty principle in quantum physics. This however provides us
a so-called ``precision paradox of chaos'', which suggests that
the prediction uncertainty of chaos is physically unavoidable, and
that even the macroscopical phenomena might be essentially
stochastic and thus could be described by probability more
economically.}


\setlength{\parindent}{0.5cm}

\section{Introduction}

One of the main goals of science is to make {\em reliable}
predictions \cite{Malescio2005}.  However, Lorenz
\cite{Lorenz1963} found that a deterministic nonlinear dynamic
system might have unpredictable solutions: for example, the famous
Lorenz's equation
\begin{equation}
\left\{
\begin{array}{l}
\dot{x}(t) = \sigma \left[y(t)-x(t)\right],\\
\dot{y}(t) = R \; x(t)-y(t)-x(t)\; z(t), \\
\dot{z}(t) = x(t)\; y(t) + b \; z(t),
\end{array}
\right.\label{geq:lorenz:original}
\end{equation}
where $\sigma, R$ and $b$ are physical parameters, and the dot
denotes the differentiation with respect to the time $t$,
respectively, has ``nonperiodic'' solutions in many cases such as
$\sigma=10, R = 28, b =-8/3$, which were named ``chaos''  later by
Li and Yorke \cite{TYLI1975}.  Chaos is a feature in all sciences
\cite{Hanski1993,Ashwin2003} and has the famous ``butter-fly
effect'': solutions are  exponentially sensitive to initial
conditions and thus a tiny variation of initial conditions may
bring huge difference of numerical results for a long time $t$.

Mostly, nonlinear continuous-time dynamical systems are
investigated by means of numerical integration algorithms
\cite{Parker1989}, which model a continuous-time system by a
discrete-time system.  Numerical simulations are widely applied to
study chaos, and such kind of computations are often called
``numerical experiments''.  Unfortunately,  numerical errors are
inherent in any numerical algorithms: there always exist the
so-called ``numerical noise'', i.e. the round-off error and
truncation error.  For evaluating floating-point expressions, the
magnitude of the round-off error depends upon the hardware used.
Typically, double-precision representations use 64 bits and are
accurate to  16 decimal places.  The truncation error is
introduced when an infinite series is truncated to a finite number
of terms. The local round-off error and truncation error propagate
together in a rather complicated way, which cause the so-called
global round-off error and global truncation error
\cite{Parker1989}. So, like physical experiments, numerical
experiments are also {\em not} perfect.

Exponential sensitivity to initial conditions implies that an
arbitrarily small {\em local} error greatly affects the
macroscopic behavior of a nonlinear dynamical system with chaos,
no matter whether such local error comes from the initial
condition (due to the inaccuracy of measured input data) or from
the ``numerical noise'' mentioned above. So, not only the
inaccuracy of initial conditions but also both of the round-off
error and truncation error at {\em each} time-step eventually
affect  the ``long-time'' behavior of a chaotic dynamical system.
Thus, theoretically speaking, all results of chaos given by
``numerical experiments'' are a kind of admixture of ``pure''
solutions of nonlinear dynamical systems and rather complicated
propagations of the round-off error, the truncation error, and the
inaccuracy of initial data. Note that a lots of conclusions about
chaos are based on such kind of {\em inaccurate} computed data,
although it has been mathematically proved that Lorenz attractor
indeed exists \cite{Stewart2000}. Are these conclusions based on
``impure'' chaotic time series believable? Are they different from
those given by the ``pure'' chaotic solutions (with negligible
numerical noise) if such kind of ``pure'' solutions exist?
Obviously, if the answers to these questions are negative, our
knowledge about chaos must be changed completely.

A system of continuous-time differential equations may have
various discrete-time difference approximations with different
time-step $\tau$.  Each of them has different dynamic properties.
It has been found  \cite{Cloutman1996,Cloutman1998} that computed
results given by some discrete-time difference schemes are
parasitic, which have no physical meanings at all. For example,
when the exact time-dependent solution of a set of nonlinear
differential equations is known to be periodic, there is sometimes
a range of the time-step $\tau$ where the computed solution of the
finite difference equations is chaotic
\cite{Cloutman1998,Lorenz1989}.  This kind of nonphysical
parasitic solutions is called computational chaos (CC)
\cite{Lorenz1989}. By contraries, when the exact solution is known
to be chaotic, computed solutions are however periodic within a
range of time-step $\tau$, and this numerical phenomenon is called
computational periodicity (CP) \cite{Lorenz2006}. So, {\em
computed} dynamic behaviors observed for a finite time-step in
some nonlinear discrete-time difference equations sometimes have
nothing to do with the {\em exact} solution of the original
continuous-time differential equations at all, as pointed out by
many researchers
\cite{Cloutman1996,Cloutman1998,Lorenz1989,Lorenz2006,Teixeira2007}.

Lorenz \cite{Lorenz2006} investigated the influence of the
time-step $\tau$ on the long-term dynamic properties of a system
of three nonlinear differential equations
\begin{equation}
\left\{
\begin{array}{l}
\dot{x}(t) = -y^2(t)-z^2(t)-A x(t)+ A F,\\
\dot{y}(t) = x(t) y(t) -B x(t) z(t)-y(t) + G, \\
\dot{z}(t) = B x(t) y(t)+x(t) z(t)-z(t),
\end{array}
\right.\label{geq:lorenz:II}
\end{equation}
where $A, B, F$ and $G$ are constant physical parameters.  Using a
numerical procedure based on the $M$th-order truncated Taylor
series in the interval $t\in[t_n,t_n+\tau]$:
\begin{eqnarray}
\left\{
\begin{array}{lcl}
x(t) &=& x(t_n)+ \sum\limits_{k=1}^{M}  \alpha_k \; (t-t_n)^k, \\
y(t) &=& y(t_n)+ \sum\limits_{k=1}^{M} \beta_k  \; (t-t_n)^k,\\
z(t) &=& z(t_n)+ \sum\limits_{k=1}^{M} \gamma_k \; (t-t_n)^k,
\end{array}
\right.  \label{Taylor-procedure}
\end{eqnarray}
where
\[   \alpha_k = \frac{1}{k!} \frac{d^k
x(t_n)}{d t^k}, \;\; \beta_k = \frac{1}{k!} \frac{d^k y(t_n)}{d
t^k}, \;\; \gamma_k = \frac{1}{k!} \frac{d^k z(t_n)}{d t^k},   \]
Lorenz \cite{Lorenz2006} studied the relationship between
computational periodicity (CP) and the time-step $\tau$, the order
$M$ and so on.  It is commonly believed that Eq.
(\ref{geq:lorenz:II}) has chaotic solution when $A=1/4, B=4, F=8$
and $G=1$.  However, when $M=1$, the leading Lyapnuv exponent
$\lambda_1$ changes sign frequently in the range $0<\tau <
\tau^*$, so that alternations between chaos ($\lambda_1
>0 $) and computational periodicity ($\lambda_1 <0$) occur
frequently.  Here, $\tau^*$ is the lowest value of time step above
which the computational instability (CI) occurs.  As one
continuously decreases the time-step $\tau$, chaos is first
observed in the range $0.0402 \leq \tau \leq 0.0435 $, then
disappears, and is observed again in the range $0.0344 \leq \tau
\leq 0.0374$, then disappears once more for the smaller $\tau$,
and is observed again when $\tau=0.0028$, but disappears once
again for the smaller $\tau$ until $\tau=0.00039$. Rather
unexpectedly, even the different chaotic solutions in case of
$M=1$ have unlike features: the intersections with plane $z=0$ for
the attractor with $\tau=0.037$ and $\tau=0.042$ are quite
dissimilar.  Similar numerical phenomena are observed for
different physical parameters.  When $M=2$ or 3, the range of
$\tau$ where the computational periodicity occurs is still much
larger than the range where the true chaos is captured. Even when
$M=4$ the ranges are nearly the same.  Only when $M=6$ does the
computational periodicity almost disappear.  For details, please
refer to Lorenz \cite{Lorenz2006}.

Currently, Teixeira {\em et al} \cite{Teixeira2007} investigated
the time-step sensitivity of three nonlinear atmospheric models of
different level of complexity, i.e. Lorenz equations
(\ref{geq:lorenz:original}), a quasigeostrophic (QG) model and a
global weather prediction system (NOGAPS).  They illustrated that
numerical convergence can not be guaranteed forever for fully
chaotic systems, because the critical time of decoupling of
numerical chaotic solutions for different time step follows a
logarithmic rule as a function of time step for the three models.
Besides, for regimes that are not fully chaotic, different time
step may lead to different model climates and even different
regimes of the solution.   For instance, for Lorenz equation
(\ref{geq:lorenz:original}) with fully chaotic solution in case of
$\sigma=10, R = 28, b =-8/3$, Teixeira {\em et al}
\cite{Teixeira2007} employed the same second-order numerical
scheme as used by Lorenz \cite{Lorenz1963} in 1963, with three
different time steps: $\tau = 0.01$ (used by Lorenz
\cite{Lorenz1963}), $\tau=0.001$ and $\tau = 0.0001$
nondimensional Lorenz time units (LTU).  All solutions are quite
close to each other for some initial time.  Then, the solution
with $\tau=0.01$ LTU decouples at about 5 LTU from the other two
solutions, and the solution with $\tau=0.001$ LTU decouples at
about 10 LTU from the solution with $\tau=0.0001$ LTU.  It is
interesting that all of these three solutions agree well in the
interval $0 \leq t \leq 5$ LTU.  Besides, Teixeira {\em et al}
\cite{Teixeira2007} found that the decoupling time $T_c$ follows
approximately $T_c = \alpha - \beta \log_{10} \tau$, where
$\alpha>0$ and $\beta>0$ are constants.  Replacing $\tau$ by
$\tau^N$ in this formula, where $N$ is the order of the numerical
scheme, Teixeira {\em et al} \cite{Teixeira2007} deduced the
conclusion that $T_c$ should be directly proportional to $N$,
although no numerical proofs are directly given to support it.
They showed that, in case of $\sigma=10, b =-8/3$ and $R = 19$,
the solution of $x(t)$ with $\tau=0.01$ LTU converges to a stable
positive fixed-point, while the other two solutions with
$\tau=0.001$ LTU and $\tau=0.0001$ LTU converge to a stable
negative fixed-point.  Besides, for Lorenz equation without fully
chaotic behavior in case of $\sigma=10, b =-8/3$ and $R = 21.3$,
the solutions of $x(t)$ with $\tau=0.01$ LTU and $\tau=0.0001$ LTU
converge to a stable fixed point, but the solution with
$\tau=0.001$ LTU keeps chaotic. Thus, different time-steps may
lead to not only the uncertainty in prediction but also
fundamentally different regimes of the solution.  The solutions of
$y(t)$ and $z(t)$ behave similarly. By means of the forth-order
Runge-Kutta scheme,  the same general findings mentioned above are
confirmed. For details, please refer to Teixeira {\em et al}
\cite{Teixeira2007}.

Facing these numerical phenomena mentioned above, one might be
confused: how can we ensure that a computed solution with chaotic
behavior is {\em indeed} chaotic but not a so-called computational
chaos (CC), and that a computed long-term solution with
periodicity is {\em indeed} periodic but not a computational
periodicity (CP)? Unfortunately,  {\em exact} chaotic solutions
for nonlinear differential equations have never been reported. So,
one even has reasons to believe that ``all chaotic responses are
simply numerical noises and have nothing to do with differential
equations'' \cite{YaoHughes-Tellus-2008,YaoHughes-JAS-2008}.

These observed phenomena of the uncertainty of long-term
predictions, computational chaos (CC) and computational
periodicity (CP) reveal some fundamental features of nonlinear
differential equations with chaos.   Obviously, both
 CC and CP are parasitic solutions and have no physical meanings at
 all, and thus should be avoided in numerical simulations.   It
 seems that chaotic
 numerical results are made of reliable and unreliable
 data.  And different numerical schemes
 might lead to completely different predictions, as pointed out by
 Lorenz  \cite{Lorenz1989, Lorenz2006}
 and Teixeira {\em et al} \cite{Teixeira2007}.   Certainly, all conclusions based on unreliable computed
 results are doubtable.  So, some fundamental concepts and general methods  should be developed to detect the reliable
 part from a given computed solution,  which is even more important than putting forward a new numerical
 scheme for nonlinear differential equations.

 This paper is organized as follows. In \S 2, a new concept, namely the critical
 predictable time $T_c$, is introduced to detect the reliable numerical solution
 from calculated chaotic results.
 In \S 3, the influence of the round-off error, the truncation
 error and the inaccuracy of initial condition on the critical
 predictable time $T_c$ is investigated by using Lorenz equation as an example.
  In \S 4, some examples are given
to illustrate how computational uncertainty of prediction,
computational chaos and computational periodicity of complicated
nonlinear dynamic systems can be avoided by means of the concept
of the critical predictable time. In \S 5, the origin of
prediction uncertainty of chaos is investigated.  In \S 6, some
discussions are given.

\section{A strategy to detect reliable numerical results}

As pointed out by Yao and Hughes \cite{YaoHughes-Tellus-2008}, it
would be an exciting contribution if {\em convergent}
computational  chaotic solutions of nonlinear differential
equations could be obtained. Unfortunately, such {\em convergent}
 solutions of chaos have never  been reported.  It is even
unknown whether such kind of {\em convergent} solutions (in
traditional meaning) of chaos exist or not.  Besides, it is also
{\em not} guaranteed whether or not a computed chaotic result
obtained by the smallest time-step is closest to the {\em  exact}
chaotic solution of the continuous-time differential equations
\cite{Teixeira2007, YaoHughes-JAS-2008, Teixeira2008}.  How can we
detect a reliable one from different computed chaotic results? How
can we avoid the so-called computational chaos (CC) and
computational periodicity (CP)?

Discovering the exponential sensitivity of chaos to initial
conditions, Lorenz \cite{Lorenz1963} revealed that it is
impossible to give accurate ``long-term'' prediction of a
nonlinear dynamic system with chaotic behaviors.  The current
works  of Lorenz \cite{Lorenz2006} and Teixeira {\em et al}
\cite{Teixeira2007} further revealed the sensitivity of computed
chaotic results to numerical schemes and time-steps.  All of these
current investigations confirm Lorenz's famous conclusion:
accurate ``long-term'' prediction of chaos is impossible
\cite{Lorenz1963}. This conclusion is widely accepted today by
scientific society. However, from mathematical points of view,
this famous conclusion is not very precise, because it contains an
 ambiguous word ``long-term''.  The concept of ``long''
or ``short''  is relative: a hundred year is long for everyday
life but is rather short for the evolution of the universe.  Is 10
non-dimensional LTU (Lorenz time units) or $10^5$ LTU  long enough
for Lorenz equation?  Given a computed chaotic result, it seems
that there should exist a critical time $T_c^*$ beyond which the
computed result is unreliable or inaccurate.  If the exact (or
convergent) chaotic solution could be known, it would be easy to
determine $T_c^*$ simply by comparing the computed result with the
exact ones.   Unfortunately, no exact chaotic solutions have been
reported.  It is a pity that no theories about such critical time
$T_c^*$ have been proposed, so that the conclusion ``long-term
prediction of chaos is impossible'' is not very precise.

\begin{figure}
\centering\setcaptionwidth{5in}
\includegraphics[width=4in,angle=0]{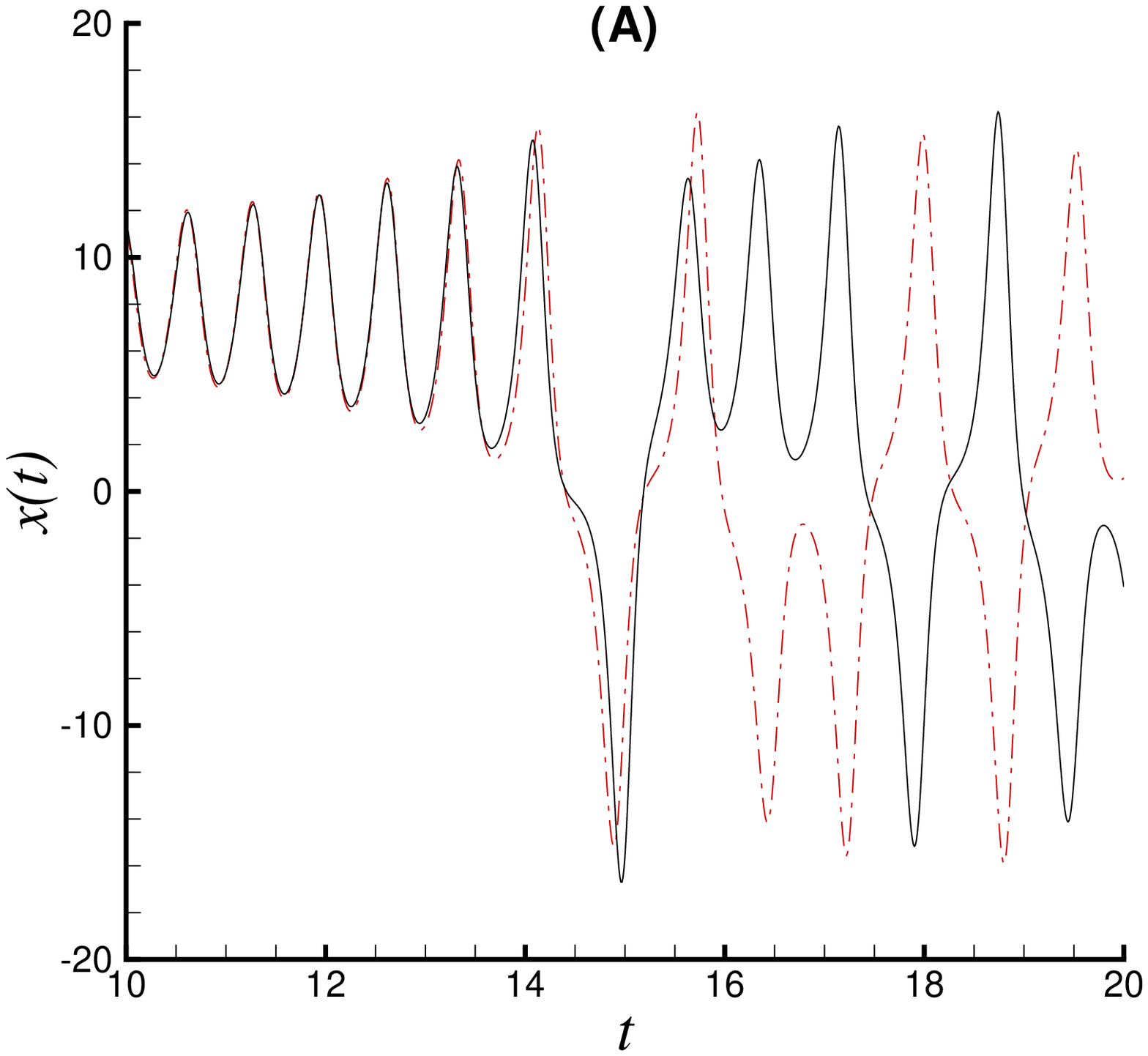}
\caption{Comparison of numerical results $x(t)$ of Lorenz's
equation by the 4th-order Runge-Kutta's method when $\sigma=10, R
= 28, b =-8/3$ and $x(0)=-15.8, y(0)=-17.48, z(0)=35.64$. Solid
line: $\tau = 10^{-5}$; Dash-dotted line: $\tau = 10^{-2}$.}
\label{figure:RK-A}

\centering\setcaptionwidth{5in}
\includegraphics[width=4in,angle=0]{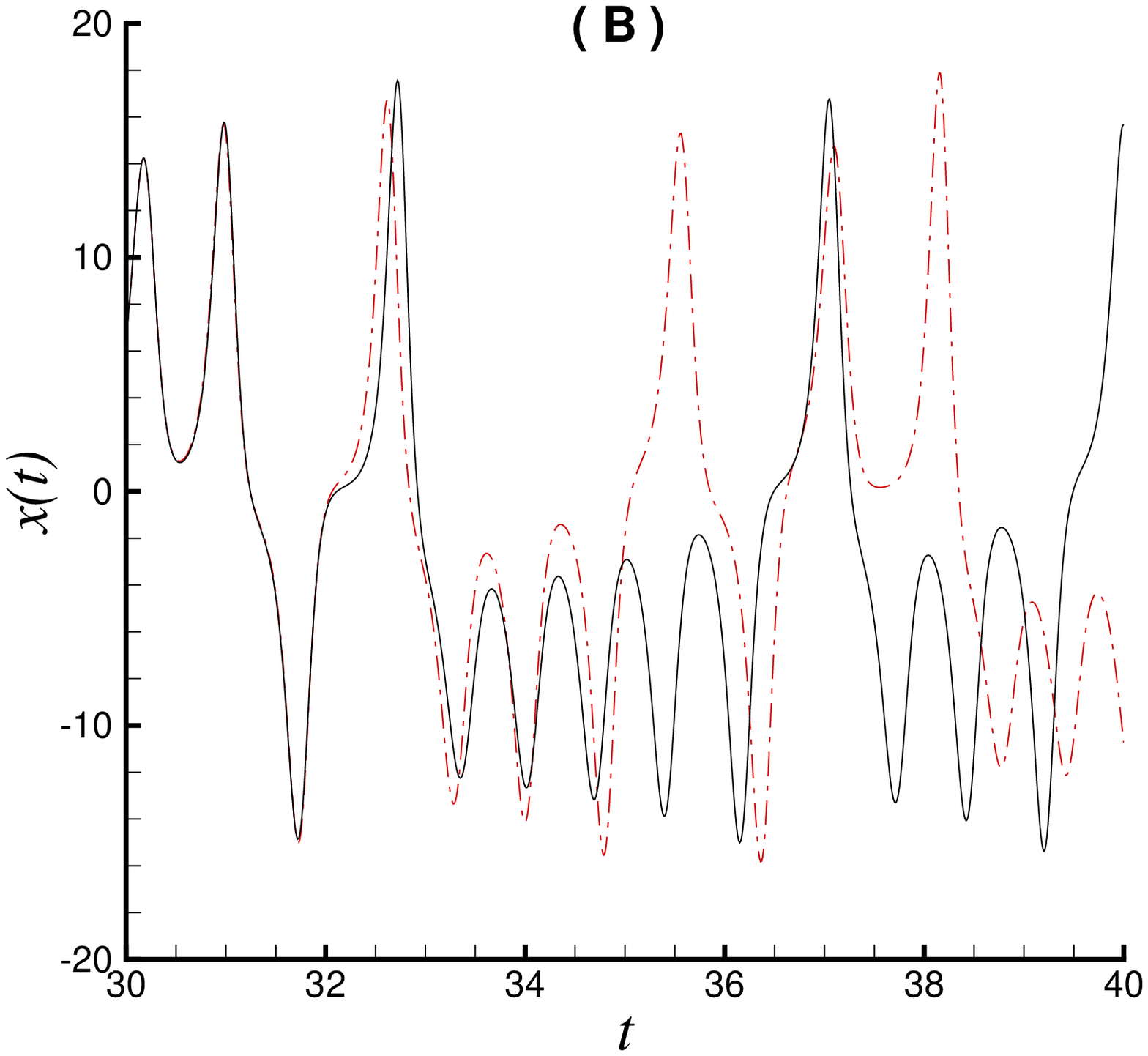}
\caption{Comparison of numerical results $x(t)$ of Lorenz's
equation by the 4th-order Runge-Kutta's method when $\sigma=10, R
= 28, b =-8/3$ and $x(0)=-15.8, y(0)=-17.48, z(0)=35.64$. Solid
line: $\tau = 10^{-5}$; Dash-dotted line: $\tau = 10^{-4}$.}
\label{figure:RK-B}
\end{figure}

Lorenz \cite{Lorenz1989, Lorenz2006} and  Teixeira {\em et al}
 \cite{Teixeira2007} confirmed such a numerical {\em fact} that two computed chaotic results given by either different
 time-step $\tau$ or different numerical schemes are rather close
 to each other from the same initial state, until they decouple at
 a critical time $T_c$, as illustrated in Figs. \ref{figure:RK-A} and
 \ref{figure:RK-B} for comparisons of numerical results of Lorenz's
equation by means of the 4th-order Runge-Kutta's method with
different time-steps in case of $\sigma=10, R = 28, b =-8/3$ and
$x(0)=-15.8, y(0)=-17.48, z(0)=35.64$.   Note that the numerical
result given by the time-step $\tau=10^{-2}$ LTU decouples the
result given by $\tau=10^{-5}$ LTU at about 14.5 LTU, as shown in
Figure \ref{figure:RK-A}, and the result given by $\tau=10^{-4}$
LTU decouples the result given by $\tau=10^{-5}$ LTU at about 33.5
LTU, as shown in Figure~\ref{figure:RK-B}, respectively.
 Besides, Parker \cite{Parker1989} also pointed out that a {\em
practical} way of judging the accuracy of numerical results of a
 nonlinear dynamical system is to use two (or more) {\em
different} routines to integrate the {\em same} system: the
initial time interval over which the two  results agree is then
{\em assumed} to be accurate and predictable. More precisely
speaking, the computed results beyond the critical decoupling time
 $T_c$ are not reliable.  Here, it should be emphasized that, up to now, it is even {\em not} guaranteed
 that the computed results in the {\em whole} region $0\leq t < T_c$ given by two different
 time-steps $\tau$ or various numerical schemes
 are convergent or very close to the ``exact'' solution, especially
 when the time-steps are very close or the numerical schemes are rather similar.   Even so,  we have
 many reasons to {\em assume} that the computed chaotic results are reliable in the region $0\leq t <
 T_c$, if we properly choose two (or more \footnote{Obviously, it is better to compare computed  results
  given by  disparate numerical schemes with different time-steps: the more, the better.
 }) different time-steps and/or numerical schemes.   This is mainly because the computed results in the interval $0 \leq t \leq T_c$ are ``predictable'':
 one can get nearly the same results by different time-step and/or disparate numerical schemes.  In this
way, we define a {\em time-scale} for the concept
 ``long-term'' of a computed chaotic result: $t$ is regarded to be ``long-term'' if $t > T_c$.
 According to Lorenz \cite{Lorenz1963, Lorenz1989, Lorenz2006} and  Teixeira {\em et al}
 \cite{Teixeira2007},  $T_c$ is sensitive to initial condition, time-step and
  numerical schemes used to compute the {\em two} different results of the same nonlinear dynamic
 system with chaos.   For convenience, we call $T_c$
the ``critical predictable time''.   Obviously, $T_c$ is dependent
upon nonlinear differential equations, time-step and numerical
schemes, thus the so-called ``long-term'' is also a relative
concept.

The so-called ``critical predictable time'' $T_c$ can be defined
in different ways.  According to Teixeira {\em et al}
 \cite{Teixeira2007}, a numerical result given by the smallest time-step is {\em assumed} to be closest to the
 exact
 solution.   So, Teixeira {\em et al}
 \cite{Teixeira2007} defined $T_c$  by means of the state vector L2 norm error between
 the result obtained by the smallest time-step and the result by a larger
 one.  This kind of definition includes the error at each
 time-step and thus is a global one for decoupling.  However, decoupling of two curves is essentially a local occurrence.
 Thus,  we give here a local
 definition of ``critical predictable time'' $T_c$, which is based on geometrical characteristic of decoupling
 of two curves and thus is straightforward.    Mathematically,  let $u_1(t)$
and $u_2(t)$ denote two time series given by different numerical
routines for a given dynamical system. The so-called ``critical
predictable time'' $T_c$ for $u_1(t)$ and $u_2(t)$ is determined
by the criteria
\begin{equation}
\dot{u}_1 \;  \dot{u}_2 < -\epsilon, \;\;\;
\left|1-\frac{u_1}{u_2}\right|> \delta, \;\;\; \mbox{at $t = T_c$,
}\label{maximum-time-criteria}
\end{equation}
where $\epsilon >0$ and $\delta >0$ are two small constants (we
use  $\epsilon=1$ and $\delta=5\%$ in this paper). Mathematically,
the critical predictable time $T_c$ can be interpreted as follows:
the influence of truncation error, round-off error and inaccuracy
of initial condition on numerical solutions is negligible in the
interval $0 \leq t \leq T_c$, so that the computed result is
predictable and thus can be regarded as a reliable solution in
this interval.  Using the concept of the critical predictable time
$T_c$, the famous statement that ``accurate long-term prediction
of chaos is impossible'' can be more precisely expressed as that
``accurate prediction of chaos beyond the critical predictable
time $T_c$ is impossible''.  Here, $T_c$ is regarded as a critical
point:  computed results beyond the critical predictable time
$T_c$ are doubtable and unreliable. Thus, the critical predictable
time $T_c$ provides us a strategy to detect the reliable part from
a given numerical result.

As pointed out by Lorenz \cite{Lorenz1989,Lorenz2006},
computational chaos (CC) and  computational periodicity (CP) are
mainly based on the evaluation of Lyapnunov exponent, which is a
long-term property. As mentioned above, any computed results for
$t
> T_c$ are doubtable and unreliable, and thus have no meanings.
Unfortunately, most of  computed  ``long-term'' solutions  are
often far beyond the critical predictable time  $T_c$, and thus
all related  conclusions or computations based on these
``long-term'' numerical results, such as computational chaos,
computational periodicity, Lyapunov exponent and attractors, are
doubtable and unreliable, too.  Note that, using the concept of
the critical predictable time $T_c$, the 3rd figure given by
Teixeira {\em et al} \cite{Teixeira2007} should be interpreted in
such a new way: the critical predictable time $T_c$ for three
computed results given respectively by $\tau = 0.01, 0.001$ and
$0.0001$ LTU is less than 15 LTU, so all computed results beyond
$t > 15$ LTU have no meanings, and thus one can {\em not} make
such a conclusion that ``different time-step may not only lead to
uncertainty in the predictions after some time, but also lead to
fundamentally different regimes of the solution''
\cite{Teixeira2007}.  In fact, by means of the concept of the
critical predictable time, the computational uncertainty of
prediction, computational chaos (CC) and computational periodicity
(CP) can be avoided completely, as shown in \S 4 for details.

As suggested by Parker \cite{Parker1989},  all numerical results
should be interpreted properly.  The critical predictable time
$T_c$ can be understood as follows:  the influence of  truncation
error, round-off error and inaccuracy of initial condition on
computed chaotic solutions is almost negligible in the time
interval $0 \leq t \leq T_c$.  Thus, the so-called critical
predictable time $T_c$ provides us a scale to investigate chaos in
a more precise way.  This new concept may greatly deepen and
enrich our understanding about chaos, as shown later.

\section{Influence of numerical scheme and data inaccuracy on $T_c$}

 Since computed
chaotic solutions beyond the critical predictable time $T_c$ are
unreliable, a numerical result with small $T_c$ is almost useless.
Thus, it is necessary to obtain reliable chaotic solutions with
large enough $T_c$.

Without loss of generality, let us consider Lorenz's equation
(\ref{geq:lorenz:original}) in case of $\sigma=10, R = 28, b
=-8/3$ with the exact initial condition $ x(0)=-15.8$,
$y(0)=-17.48$, $z(0)=35.64$.  Using the 4th-order Runge-Kutta's
method with different time increment
 $\tau = 10^{-2}$,  $10^{-3}$, $10^{-4}$ and
$10^{-5}$ LTU (Lorenz time units), the corresponding critical
predictable times $T_c$ of computed chaotic results are about 13.7
LTU, 24.5 LTU and 32.6 LTU, respectively, as shown in Figs.
\ref{figure:RK-A} and \ref{figure:RK-B}.   So, by means of
traditional numerical methods, the critical predictable time $T_c$
of computed chaotic results is often not long enough.  Teixeira
{\em et al} \cite{Teixeira2007} found that, for given numerical
scheme, the time of decoupling of numerical chaotic solutions for
different time step follows a logarithmic rule as a function of
time step.  Thus, the time-step should be exponentially small for
a give critical predictable time $T_c$.  Lorenz \cite{Lorenz2006}
reported some qualitative influences of numerical schemes (based
on truncated Taylor series at a few different orders $M$) on the
computed chaotic results, but he did not give a quantitative
relationship between the critical predictable time $T_c$ and the
approximation order $M$.  Besides, it is a pity that the influence
of round-off error on the time of decoupling of computed chaotic
results given by different numerical schemes is neglected, mainly
because traditional digital computers use in general the
floating-point data in either single or double precision only.

Currently, some advanced symbolic computation software, such as
Mathematica and Maple, are widely used.  In this paper, the
symbolic computation software Mathematica is employed for the
first time as a computational tool to investigate the influence of
various numerical schemes based on truncated Taylor series
(\ref{Taylor-procedure}), round-off error and inaccuracy of
initial condition on the critical predictable time $T_c$. From the
view-point of round-off error, symbolic computation is completely
different from evaluating floating-point expressions: the
round-off error can be almost neglected or even {\em avoided} by
means of symbolic computation. For example, by means of symbolic
computation, we can have the {\em exact} result $ 1/2 + 1/3 =5/6$.
Note that, using numerical computation with double precision
representations, one has only the {\em approximate} result
$1/2+1/3 \approx 0.8333333333333333$, whose round-off error is
about $10^{-16}$. Besides, using the Mathematica command {\bf
N[Pi, 800]}, we can get the approximation $\pi\approx
3.1415926535897932384626433832 \cdots$, which is accurate even to
800 decimal places!   Using such precise data representation based
on symbolic computation software, the round-off error can be
almost neglected.   Let $K$ denote the number of decimal places of
all data used in the symbolic software in this paper.  By means of
different values of $K$, it is easy to investigate the influence
of the round-off error on $T_c$, as shown later. Furthermore, by
means of the scheme (\ref{Taylor-procedure}), the system of Lorenz
equations (\ref{geq:lorenz:original}) is approximated by a
time-continuous system in each interval $t\in[t_n, t_n + \tau]$ as
the truncated $M$th-order Taylor's expansion.  Obviously, the
truncation error of this scheme is determined by $M$. Therefore,
using symbolic computation and the analytic approach described
above, it is convenient to control the magnitude of the
truncation-error and the round-off error by means of $M$ and $K$,
respectively. Clearly, the larger the values of $M$ and $K$, the
smaller the truncation error and round-off error, respectively.
Thus, the symbolic computation software provides us a useful tool
to investigate the influence of truncation-error, round-off error
and inaccuracy of initial conditions on the critical predictable
time $T_c$.

\begin{figure}
\centering\setcaptionwidth{5.0in}
\includegraphics[width=3.5in,angle=0]{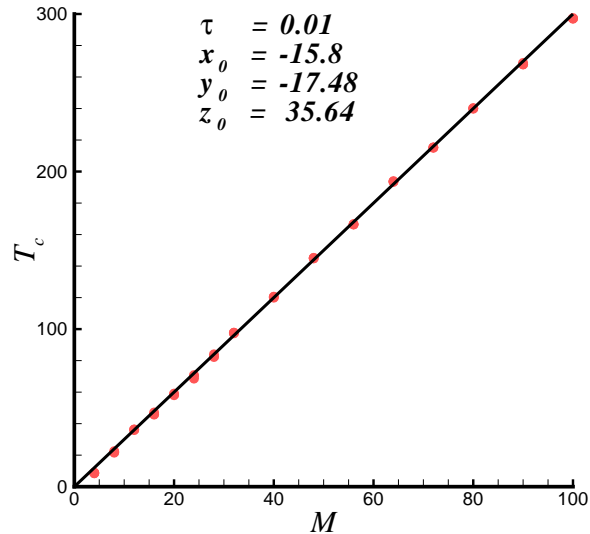}
\caption{The critical predictable time $T_c$ versus the order $M$
of truncated Taylor series (\ref{Taylor-procedure}) in case of
$\sigma=10, R = 28, b =-8/3$ and $x(0)=-15.8, y(0)=-17.48,
z(0)=35.64$ with $K = \max \left\{200,2M \right\}$. Symbols:
computed results; Solid line: $T_c = 3 M$.} \label{figure:tmax}
\end{figure}

Without loss of generality, we consider here Lorenz equation
(\ref{geq:lorenz:original}) in case of $\sigma=10, R = 28, b
=-8/3$ with the initial condition $x(0)=-15.8, y(0)=-17.48,
z(0)=35.64$ and the time step $\tau =10^{-2}$, if not mentioned
particularly.  Note that the initial condition is assumed to be
exact. To investigate the influence of the truncation error on
computed chaotic results alone, we set a large enough number of
decimal places, i.e. $K = \max\left\{ 200,2M \right\}$, where $M$
is the order of truncated Taylor series (\ref{Taylor-procedure})
of Lorenz's equation.   In this way, the round-off error is much
smaller than the truncation error,  and thus is negligible.  Since
the initial condition is assumed to be exact, there exists the
truncation error alone, whose magnitude is determined by $M$, the
order of truncated Taylor series (\ref{Taylor-procedure}) of
Lorenz equation (\ref{geq:lorenz:original}).   Using different
values of $M$ from $M=4$ to $M=110$, we get different computed
results with different truncation errors. Using
(\ref{maximum-time-criteria}) as the criteria of decoupling of two
computed trajectories, it is easy to find the corresponding
critical predictable time $T_c$ of the numerical result given by
the smaller $M$.   It is found that the critical predictable time
$T_c$ is directly proportional to $M$, i.e.
\begin{equation}
T_c \approx 3 M, \label{tmax-M}
\end{equation}
as shown in Figure \ref{figure:tmax}.

\begin{figure}
\centering\setcaptionwidth{5in}
\includegraphics[width=3.5in,angle=0]{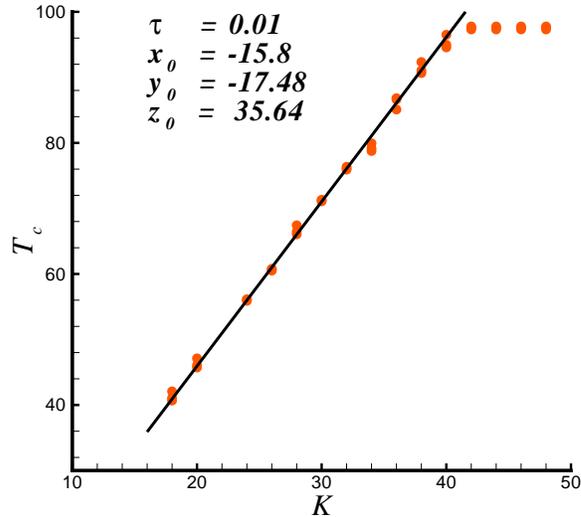}
\caption{The critical predictable time $T_c$ versus  $K$ ( the
number of accurate decimal places ) in case of $\sigma=10, R = 28,
b =-8/3$ and $x(0)=-15.8, y(0)=-17.48, z(0)=35.64$ with $M=32$.
Symbols: computed results; Solid line: $ T_c = 2.51 K - 4.26$.}
\label{figure:tmax-K}
\end{figure}

It is a little more difficult to investigate the influence of the
round-off error on chaotic solutions alone, mainly because the
round-off error might greatly increase for given $K$ when the
order $M$ is too larger than $K$.   Note that the previous formula
$T_c \approx 3 M$ (with $K = \max\left\{ 200,2M \right\}$) gives a
time interval $0\leq t \leq T_c$ in which the influence of both
truncation error and round-off error is negligible, as interpreted
before.  For example, when $M=32$, the influence of the truncation
error is negligible for $t \leq 96$.  Thus, without loss of
generality, let us consider the case of $M=32$ with different
values of $K$ ($K < 100$). Comparing the results given by
different values of $K$ ($K<100$) with the result obtained by $K =
\max\left\{ 200,2M \right\}$, we get the corresponding critical
predictable times $T_c$.  It is found that, when $K  > 40$, $T_c$
tends to the same value close to 96, respectively.   This is
mainly because, when $K$ is large enough, the round-off error is
much smaller than the truncation error.  So, the results for $K >
40$ is useless to investigate the influence of the round-off error
on $T_c$. It is also found that, when $K \leq 16$, the precision
of computation is too low relative to the $M$, the order of
approximation, so that the round-off error increases greatly.
Thus, the results with too small $K$ is also useless. So, only
results given by proper values of $K$ are useful.  It is found
that, for $18 \leq K \leq 40$, the computed critical predictable
time agree well with the formula
\begin{equation}
T_c \approx 2.51 K - 4.26,\label{tmax-K}
\end{equation}
as shown in Figure \ref{figure:tmax-K}. Furthermore, it is found
that, in general, the critical predictable time $T_c$ indeed
increases {\em linearly} with respect to $K$, the number of
accurate decimal places of results.

\begin{figure}
\centering\setcaptionwidth{5in}
\includegraphics[width=3.5in,angle=0]{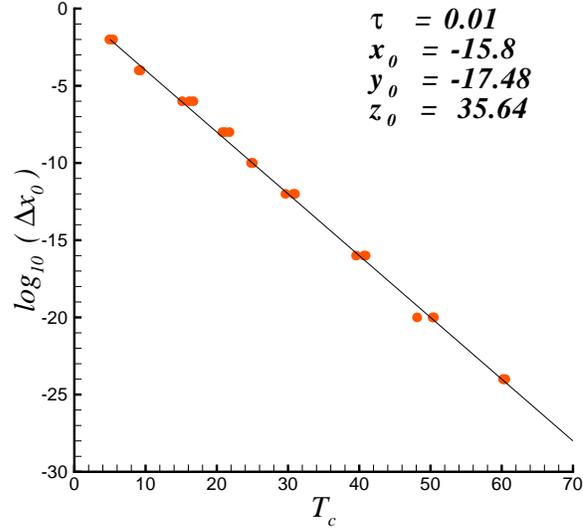}
\caption{The  critical predictable time $T_c$ versus the
inaccuracy $\Delta x_0$ of initial conditions in case of
$\sigma=10, R = 28, b =-8/3$ and $x(0)=-15.8+\Delta x_0,
y(0)=-17.48, z(0)=35.64$ with $M=100$ and $K = 200$.  Symbols:
computed results; Solid line: $log_{10}(\Delta x_0) = -0.4 T_c$.}
\label{figure:tmax-Dx0}
\end{figure}

Note that  the initial conditions are assumed to be exact in above
computations. According to our above investigations, in case of
$K=200$ and $M=100$, both of the round-off error and the
truncation error are negligible in the interval $0\leq t \leq
300$.  This provides us a convenient way to investigate the
influence of the inaccuracy of initial conditions
 on $T_c$ alone.   To do so, we add a tiny difference $\Delta x_0$ into the initial
condition $ x(0) = -15.8, y(0) = -17.48, z(0) = 35.64 $ in such a
way that
\[ x(0) = -15.8 + \Delta x_0, \]
but with the same values of $y(0)$ and $z(0)$.  Comparing the
results given by different values of $\Delta x_0 > 0$, $M=100$ and
$K=200$ with the result given by $\Delta x_0 = 0$, $M=100$ and
$K=200$, we obtain the corresponding $T_c$ by means of the
criteria (\ref{maximum-time-criteria}).   It is found that, for
given different values of $\Delta x_0$, the corresponding results
of $T_c$ agree well with the formula
\[ T_c \approx -2.5 \log_{10} (\Delta x_0), \]
as shown in Figure \ref{figure:tmax-Dx0}.  This formula can be
rewritten as
\begin{equation}
\Delta x_0 \approx 10^{ - 0.4  \;  T_c}, \label{tmax-Dx0}
\end{equation}
which means that the precision of the initial condition must
increase {\em exponentially} with respect to a given critical
predictable time $T_c$.  For example, to get a reliable chaotic
solution with $T_c = 200$ LTU, the initial condition must be with
the precision $\Delta x_0 < 10^{-80}$.  Therefore, we need a
rather precise initial condition so as to get a reliable chaotic
solution with $T_c > 200$ LTU.   Unfortunately, such precise
initial conditions are impossible in practice, as discussed in \S
5. That is exactly the reason why the ``butter-fly effect''
exists, as pointed out by Lorenz \cite{Lorenz1963}  in 1963.
However, the formula (\ref{tmax-Dx0}) might inform us much more
than the so-called ``butter-fly effect'', as discussed  in details
in \S 5.

\begin{figure}
\centering\setcaptionwidth{5in}
\includegraphics[width=3.5in,angle=0]{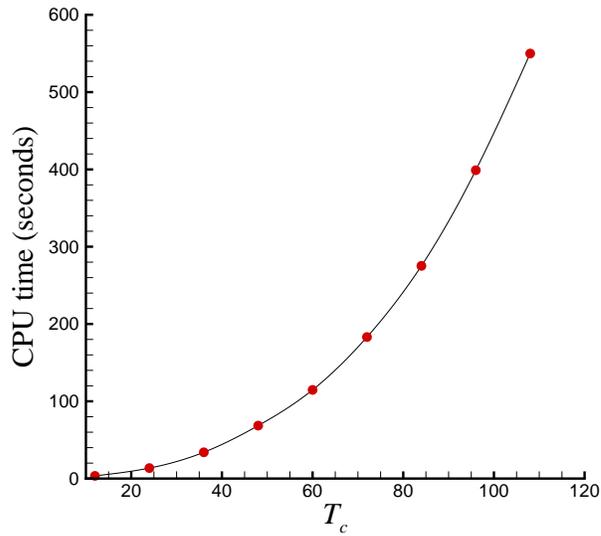}
\caption{The  used CPU time versus  $T_c$ in case of $\sigma=10, R
= 28, b =-8/3$ and $x(0)=-15.8, y(0)=-17.48, z(0)=35.64$ with $M =
T_c/3$ and $K = 200$. } \label{figure:CPU-Tc}
\end{figure}

\begin{figure}
\centering\setcaptionwidth{5in}
\includegraphics[width=3.5in,angle=0]{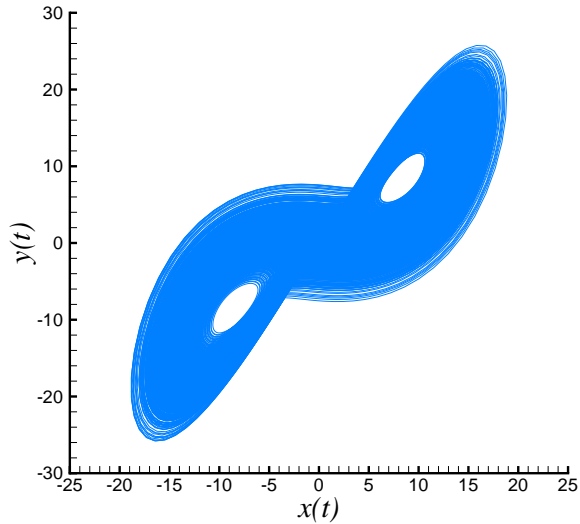}
\caption{The curve $x(t) - y(t)$ given by the reliable chaotic
result with $T_c=1200$ LTU by $\tau =0.01, M=400, K=800$ in case
of $\sigma=10, R = 28, b =-8/3$ and $x(0)=-15.8, y(0)=-17.48,
z(0)=35.64$. } \label{figure:XY}
\end{figure}

\begin{figure}
\centering\setcaptionwidth{5in}
\includegraphics[width=3.5in,angle=0]{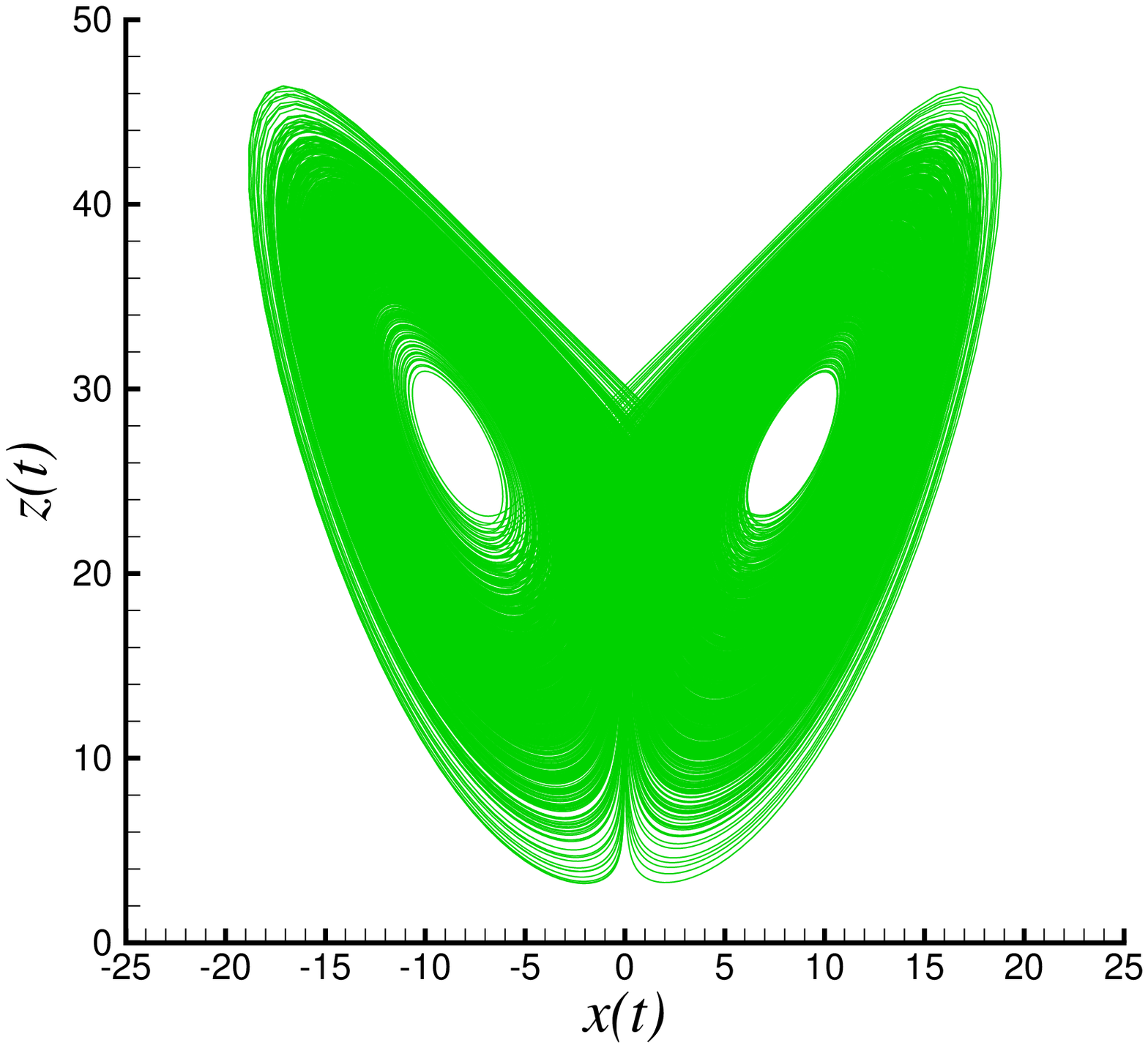}
\caption{The curve $x(t) - z(t)$ given by the reliable chaotic
result with $T_c=1200$ LTU by $\tau =0.01, M=400, K=800$ in case
of $\sigma=10, R = 28, b =-8/3$ and $x(0)=-15.8, y(0)=-17.48,
z(0)=35.64$. } \label{figure:XZ}
\end{figure}

Can we get reliable chaotic results with large enough critical
predictable time $T_c$?  Assuming that the initial condition is
exact, it is found that $T_c \approx 3 M$ generally holds in case
of $K = \max \left\{200,2M \right\}$. Therefore, $M$ (the order of
truncated Taylor series) and $K$ (the number of accurate decimal
places) should be increased linearly with respect to the critical
predictable time $T_c$.  So, theoretically speaking, for a given
$T_c$, one can always find the value of $M$ and $K$ to get a
``reliable'' chaotic solution in $0\leq t \leq T_c$.  However, the
CPU time increases with respect to $T_c$ in a power-law, as shown
in Figure \ref{figure:CPU-Tc}.  Suppose that we would like to get
a reliable chaotic solution with $T_c = 1200$ LTU.  According to
(\ref{tmax-M}), we should choose $M=400$ so as to get such a
reliable chaotic result.  In fact, by means of $M = 400$ and $K =
800$, we indeed obtain this reliable chaotic solution in the
interval $0 \leq t < 1200$ LTU, as shown in Figures
\ref{figure:XY} and \ref{figure:XZ}.  The corresponding result is
rather precise: the maximum residual error is only $1.1 \times
10^{-481}$.  However, more than 461 hours 16 minutes CPU time
(more than 19 days) is used by means of a cluster Intel Clovertown
Xeon E5310 with 8GB RAM.   To the best of our knowledge, such kind
of reliable chaotic solution of Lorenz equation in such a long
time interval has never been reported. Based on this
time-consuming computation, we are quite sure that the solution of
Lorenz equation (\ref{geq:lorenz:original}) in case of $\sigma=10,
R = 28$ and $b =-8/3$ is indeed chaotic {\em within} the interval
$0 \leq t < 1200$ LTU, as shown\footnote{The detailed data can be
downloaded via the website: http://numericaltank.sjtu.edu.cn/.} in
Table \ref{Table:XYZ}.  However, strictly speaking, it is unknown
whether or not the chaotic behavior disappears when $t>1200$ LTU.
This is because, based on our current computations, chaotic
numerical results beyond $T_c$ is unreliable. To answer this
question, one had to spend more CPU time to get a reliable chaotic
solution with even larger $T_c$.  Unfortunately, $T_c$ is always a
finite value, no matter how large it is!  And the nonlinearly
increased CPU time also indicates the impossibility to get a
reliable chaotic solution in an infinite interval $0 \leq t <
+\infty$.  This reveals from the view-point of CPU time that chaos
is unpredictable in essence.

\begin{table}
\centering\caption{Some reliable numerical results with $T_c =
1200$ LTU in case of $\sigma=10, R = 28, b =-8/3$ by means of
$M=400$, $K=800$ and $\tau=0.01$.  The detailed data can be
downloaded via the website: http://numericaltank.sjtu.edu.cn/.}
\label{Table:XYZ} \vspace{0.25cm} {\small
\begin{tabular}{|c|c|c|c|}\hline
0  &  -15.8  &  -17.48  &  35.64 \\
50  &  12.779038299490452  &  8.825054357006032  &
36.40092236534542 \\
100  &  -10.510118721506247  &  -12.17254281368225  &
27.476265630374762 \\
150  &  -1.9674157212680177  &  -2.5140404626072206  &
17.233128197642884 \\
200  &  -6.697233173381982  &  -11.911020483539128  &
13.036826414358321 \\
250  &  3.480010996527037  &  5.743865139093177  &
22.424028925951887 \\
300  &  10.197534991661733  &  3.906517722362926  &
35.33742709240441 \\
350  &  0.009240166388150502  &  -1.1520585946848019  &
20.259118270313508 \\
400  &  -1.8892476498049868  &  -3.5657880408974663  &
20.299639635504597  \\
450  &  2.3442055460290803  &  2.473910407011588  &
19.324756580383077  \\
500  &  -5.30509963157152  &  -9.425991029211517  &
12.302184230689779  \\
550  &  -9.710817000847529  &  -6.878169205988265  &
31.67393963382737 \\
600  &  -0.8635053825976141  &  0.499057856286716  &
21.581438144249077 \\
650  &  -6.249196468824656  &  -1.3133350412836564  &
30.3936296733578 \\
700  &  10.884963668216704  &  16.32989379246704  &
22.247458859587212 \\
750  &  -1.5200586402319973  &  -0.4164281272461717  &
21.530357757012936 \\
800  &  1.3963347154139534  &  2.40877126758134  &
14.590441270059282 \\
850  &  1.580132807298193  &  2.6273272210193146  &
12.83939375621528 \\
900  &  -6.449367823985297  &  -10.984642417532422  &
14.647974468278282 \\
950  &  10.098469202323805  &  0.4959032511015884  &
37.72812801085503 \\
1000  &  13.881997000862393  &  19.918303160406396  &
26.901943308376104 \\
1025  &  -2.831908677750036  &  -5.127291386139972  &
10.787422525560384 \\
1050  &  -6.0495817084397405  &  -0.5249599507390699  &
30.805747242725836 \\
1075  &  -8.445628564097573  &  -16.91583633884055  &
8.185099340204886 \\
1100  &  2.2974592711836634  &  2.299710874996516  &
19.617779431769037 \\
1125  &  -2.0420317363264457  &  -0.3357510158682992  &
23.174657463445286 \\
1150  &  -14.378782424952437  &  -11.819346602645444  &
37.319351169225996 \\
1175  &  -11.794511899005188  &  -13.181679857519981  &
29.65720151904728 \\
1200  &  2.4537546196402595  &  4.124943247158509  &
19.349201739150004 \\
 \hline
\end{tabular}
}
\end{table}

\section{Avoidance of CC, CP and computational uncertainty of prediction}

The uncertainty of prediction of chaos have two reasons. One
 is computational (or more precisely, mathematical), which is due to nonlinearity of models and the unperfect of
numerical schemes and data precision mentioned in \S 3.  The other
is physical, which is based on the fundamental physical principles
of nature.

In this section, we investigate the computational uncertainty of
chaos.  In essence, the computational uncertainty of prediction
comes from the unpredictability of trajectories, i.e. the
decoupling of different trajectories for a ``long'' time.  Using
the ``critical predictable time'' $T_c$ and regarding chaotic
results reliable only in the interval $0\leq t < T_c$, the
 numerical phenomena such as computational chaos
(CC), computational periodicity (CP) and computational uncertainty
(CU), can be avoided, as illustrated below.

\begin{figure}
\centering\setcaptionwidth{5.0in}
\includegraphics[width=4.0in,angle=0]{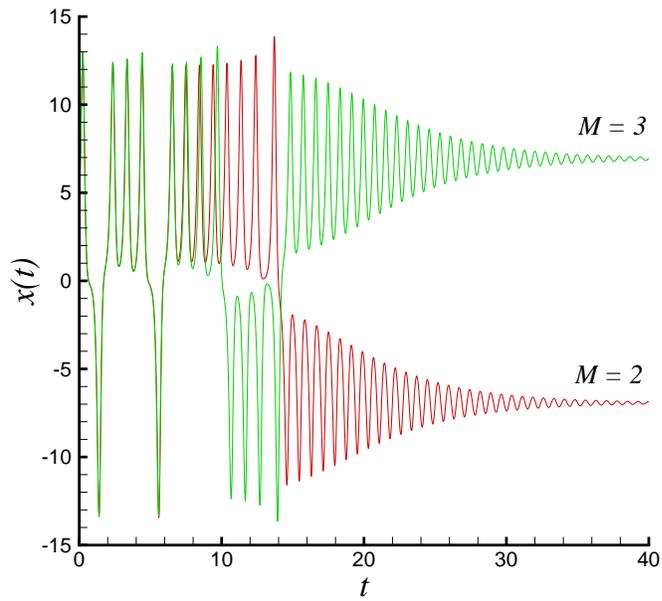}

\caption{Comparison of $x(t)$ in case of $\sigma=10, R = 19, b
=-8/3$ with the initial state $x=y=z=5$ by means of $\tau=0.01$
LTU and the $M$th-order scheme (\ref{Taylor-procedure}) based on
truncated Taylor series.  Red line: $M=2$; Green Line: $M=3$}
\label{figure:R19M2M3}
\end{figure}

\begin{figure}
\centering\setcaptionwidth{5.0in}
\includegraphics[width=4.0in,angle=0]{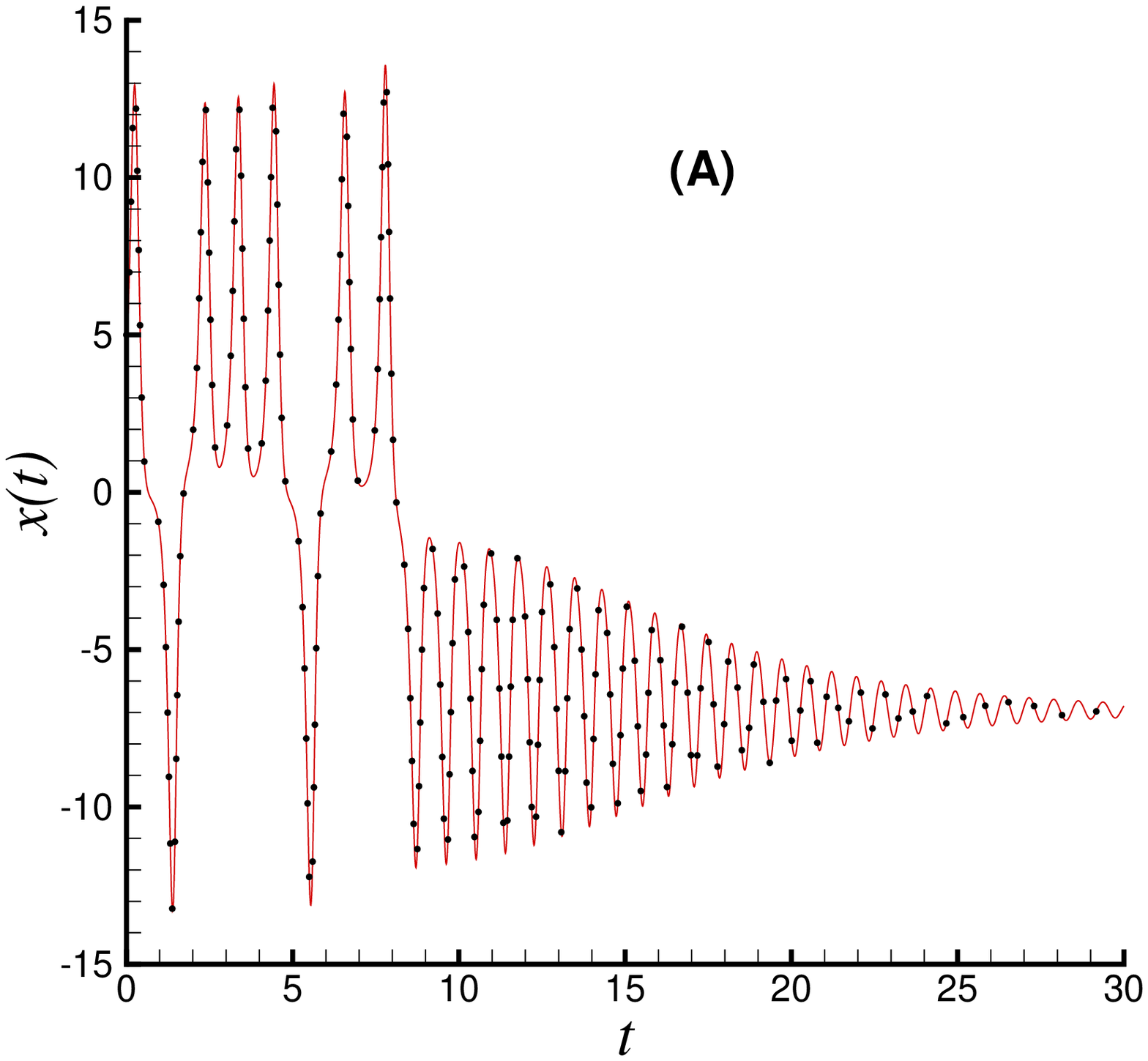}

\includegraphics[width=4.0in,angle=0]{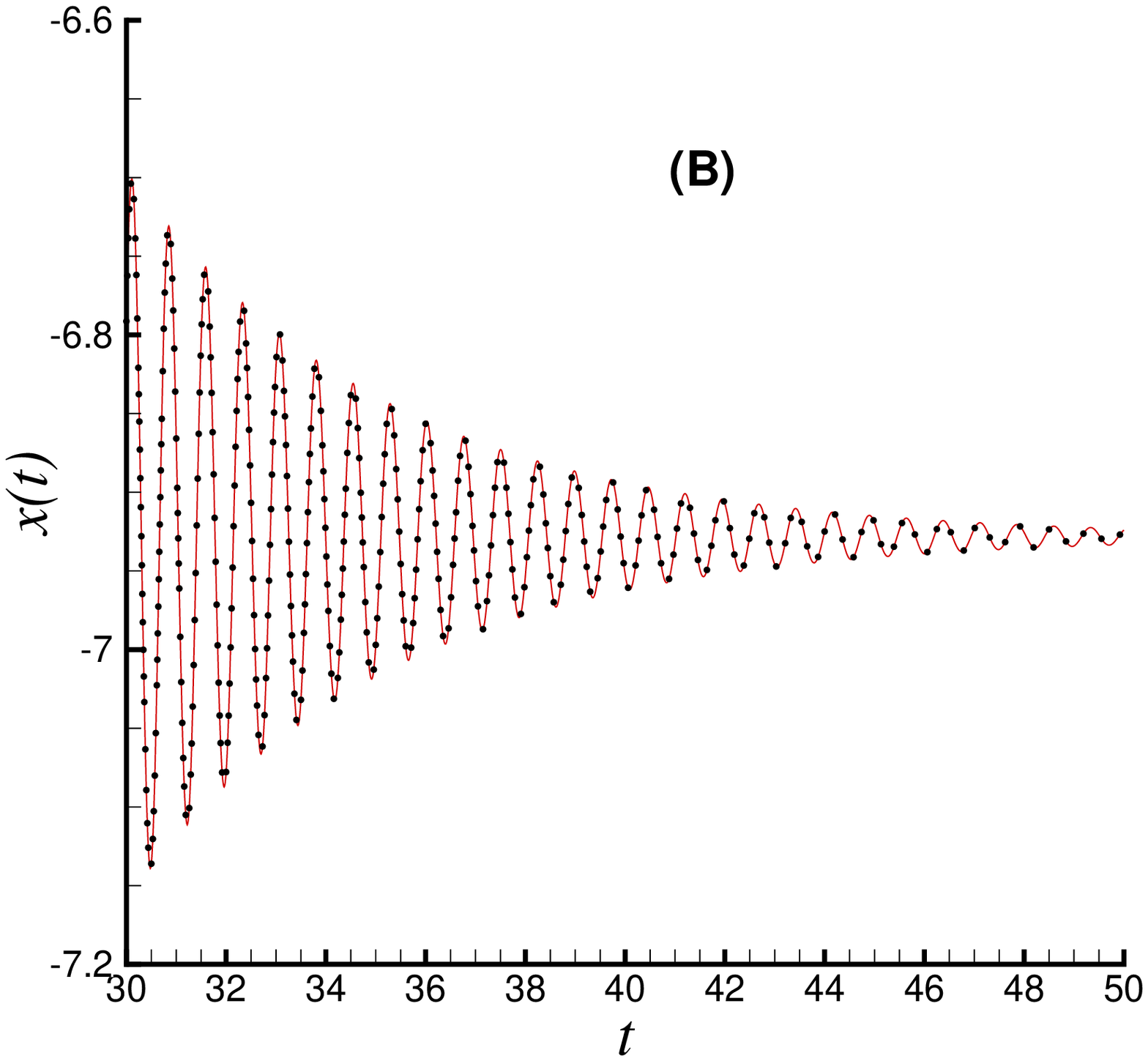}

\caption{(A) Comparison of $x(t)$ in case of $\sigma=10, R = 19, b
=-8/3$ with the initial state $x=y=z=5$ by means of $\tau=0.01$
LTU and the $M$th-order scheme (\ref{Taylor-procedure}) based on
truncated Taylor series in the interval $0 \leq t \leq 30$. Line:
$M=30$; Symbols: $M=40$. (B): the same, but in the interval $30
\leq t \leq 50$.} \label{figure:R19M30M40}
\end{figure}

It is well known that solution of Lorenz's equation
(\ref{geq:lorenz:original}) becomes unstable if $R > R_c = \sigma
(\sigma-b+3)/(\sigma+b-1)$. In case of $\sigma=10$ and $b =-8/3$,
we have the critical value $R_c =24.7368$.  Thus, in case of
$\sigma=10, b =-8/3$ and $R = 19 < R_c$ with the initial state $x
= y =z =5$, the exact solution should tend to a fixed point.
However, it is unknown which fixed point the solution tends to. It
is found that the computed result $x(t)$ given by the $M$th-order
scheme (\ref{Taylor-procedure}) with $\tau=0.01$ tends to a
negative fixed-point for $M=2$, but goes to a positive fixed point
for $M=3$, as shown in Figure \ref{figure:R19M2M3}.   Thus, at
least one of these two different predictions must be wrong.
However, based on these two computational results, it is hard to
detect which prediction is correct.  This kind of computational
uncertainty of prediction is similar to those mentioned by
Teixeira {\em et al} \cite{Teixeira2007}. Note that the critical
predictable time $T_c$ of these two computed results are only
about $T_c \approx 9$ LTU, as shown in Figure
\ref{figure:R19M2M3}, so that they are reliable only in the
interval $0 \leq t < 9$ LTU. In other words, the two computed
results in the interval $t
> 9$ are {\em unreliable} and thus has no meanings.   Therefore, based on these two
computational results, one can not give any reliable conclusions
about the fixed point.  To get a reliable prediction about the
fixed-point, we had to give a numerical result with large enough
$T_c$.  To do so, we use higher-order schemes
(\ref{Taylor-procedure}) based on the truncated Taylor series. As
shown in Figure \ref{figure:R19M30M40}, the two computed results
given by $M= 30$ and $M=40$ with $\tau =0.01$ LTU agree well in
the interval $0\leq t \leq 100$ LUT,  and both of them give the
{\em same} numerical fixed-point:
\[ x(100) = -6.928204, \;\; y(100)=-6.928204, \;\; z(100) = 18.000000.\]
 Based on these two reliable numerical results,  we are quite
sure that the exact solution $x(t)$ of Lorenz equation
(\ref{geq:lorenz:original}) tends to a negative fixed point in
case of $\sigma=10, b =-8/3$ and $R = 19 < R_c$ with the initial
state $x = y =z =5$.  In this way, the computational uncertainty
of prediction mentioned by Teixeira {\em et al}
\cite{Teixeira2007} and Lorenz \cite{Lorenz2006} can be avoided.

\begin{figure}
\centering\setcaptionwidth{5.0in}
\includegraphics[width=4.0in,angle=0]{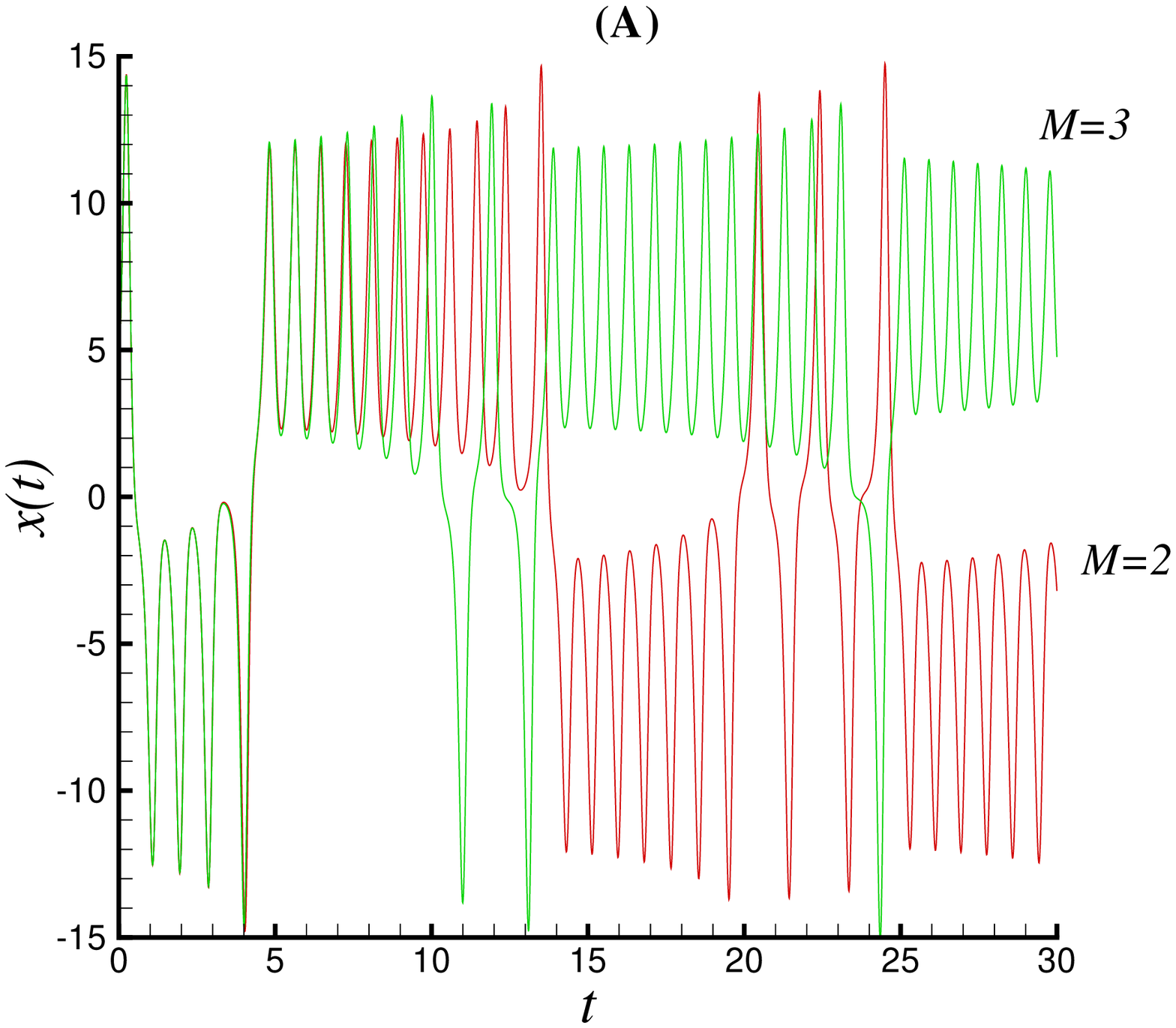}

\includegraphics[width=4.0in,angle=0]{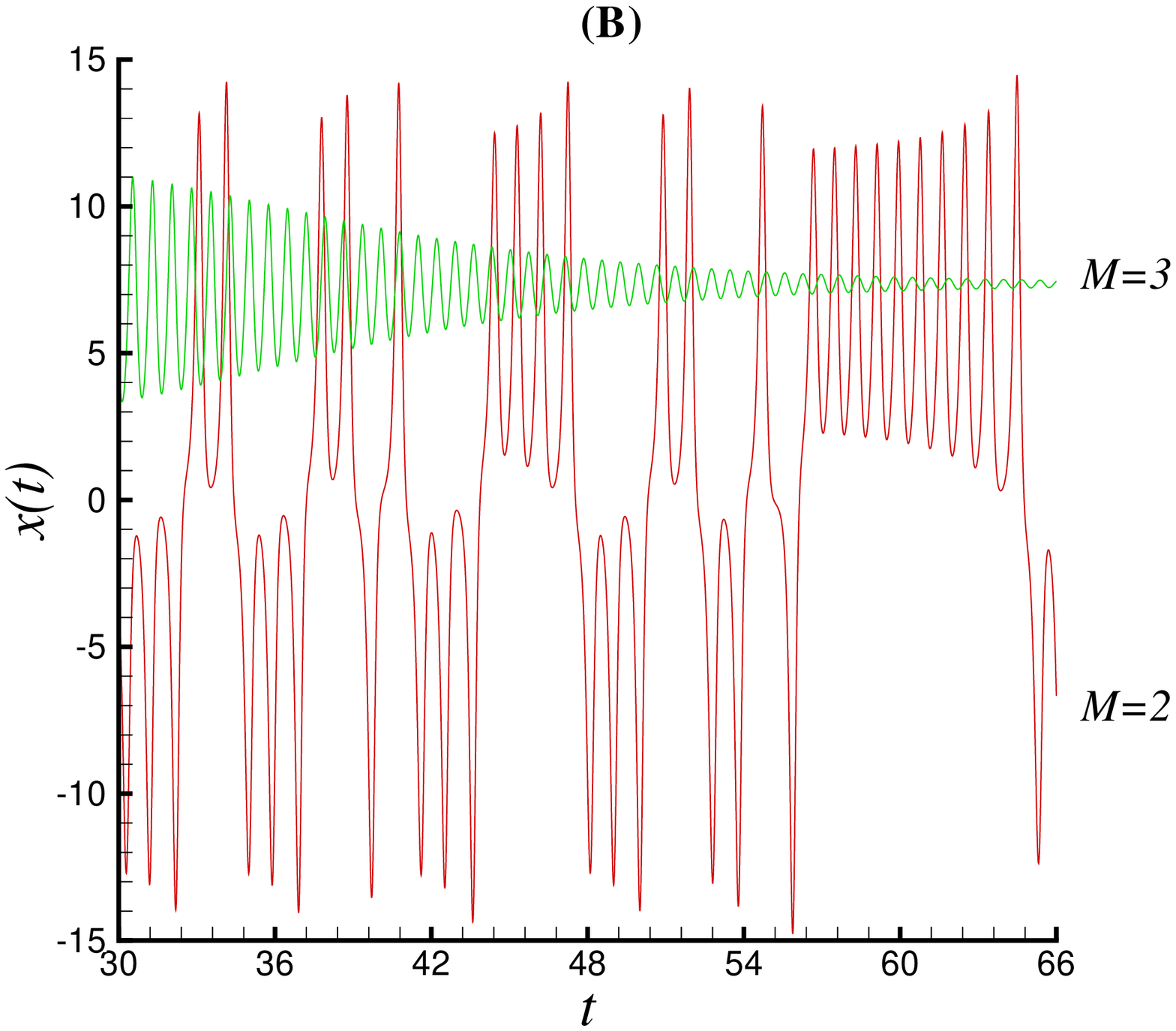}

\caption{(A) Comparison of $x(t)$ in case of $\sigma=10, R = 21.5,
b =-8/3$ with the initial state $x=y=z=5$ by means of $\tau=0.01$
LTU and the $M$th-order scheme (\ref{Taylor-procedure}) based on
truncated Taylor series in the interval $0 \leq t \leq 30$. Red
line: $M=2$; Green line: $M=3$. (B): the same, but in the interval
$30 \leq t \leq 66$.} \label{figure:R21d5M2M3}
\end{figure}

\begin{figure}
\centering\setcaptionwidth{5.0in}
\includegraphics[width=4.0in,angle=0]{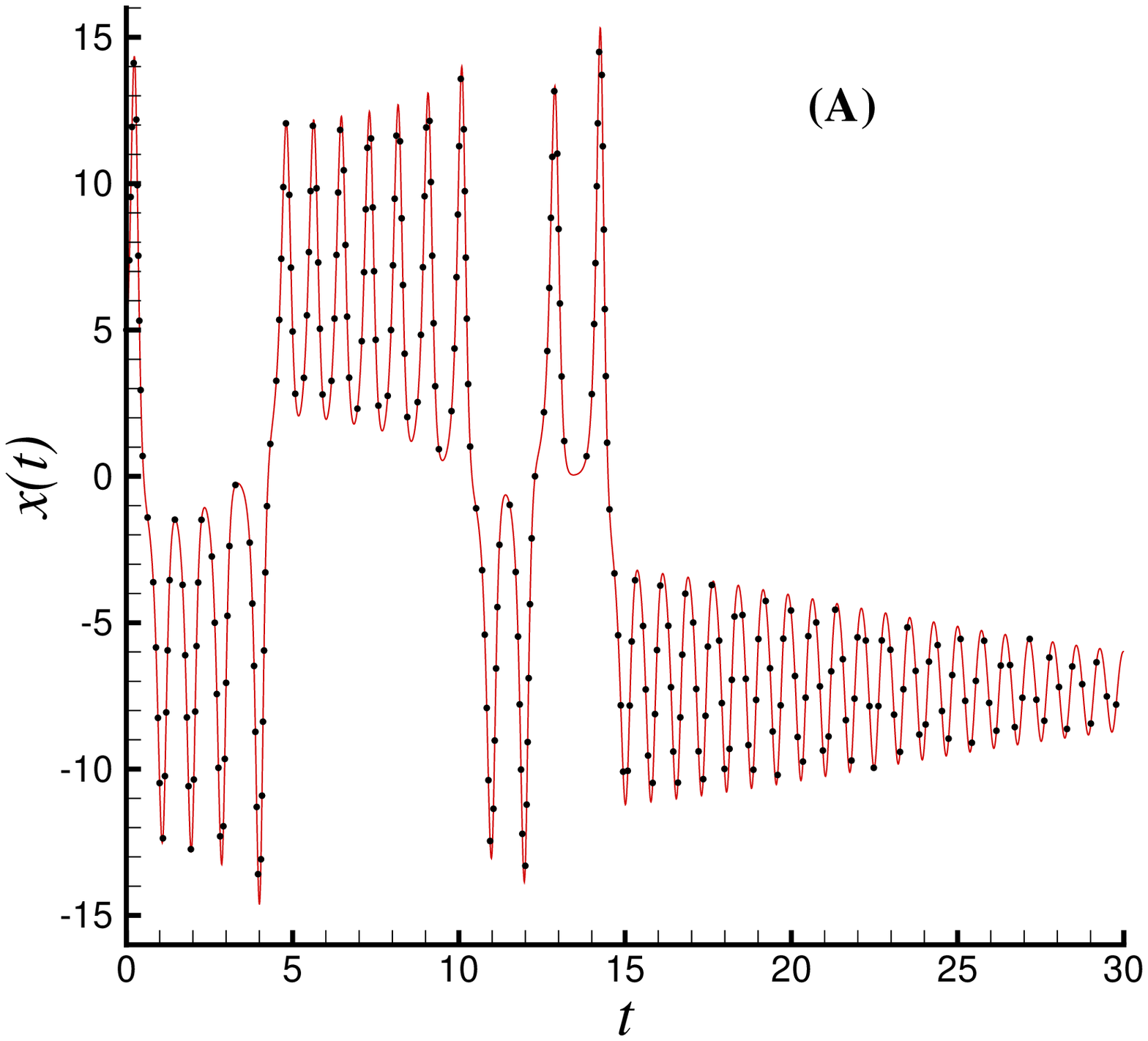}

\includegraphics[width=4.0in,angle=0]{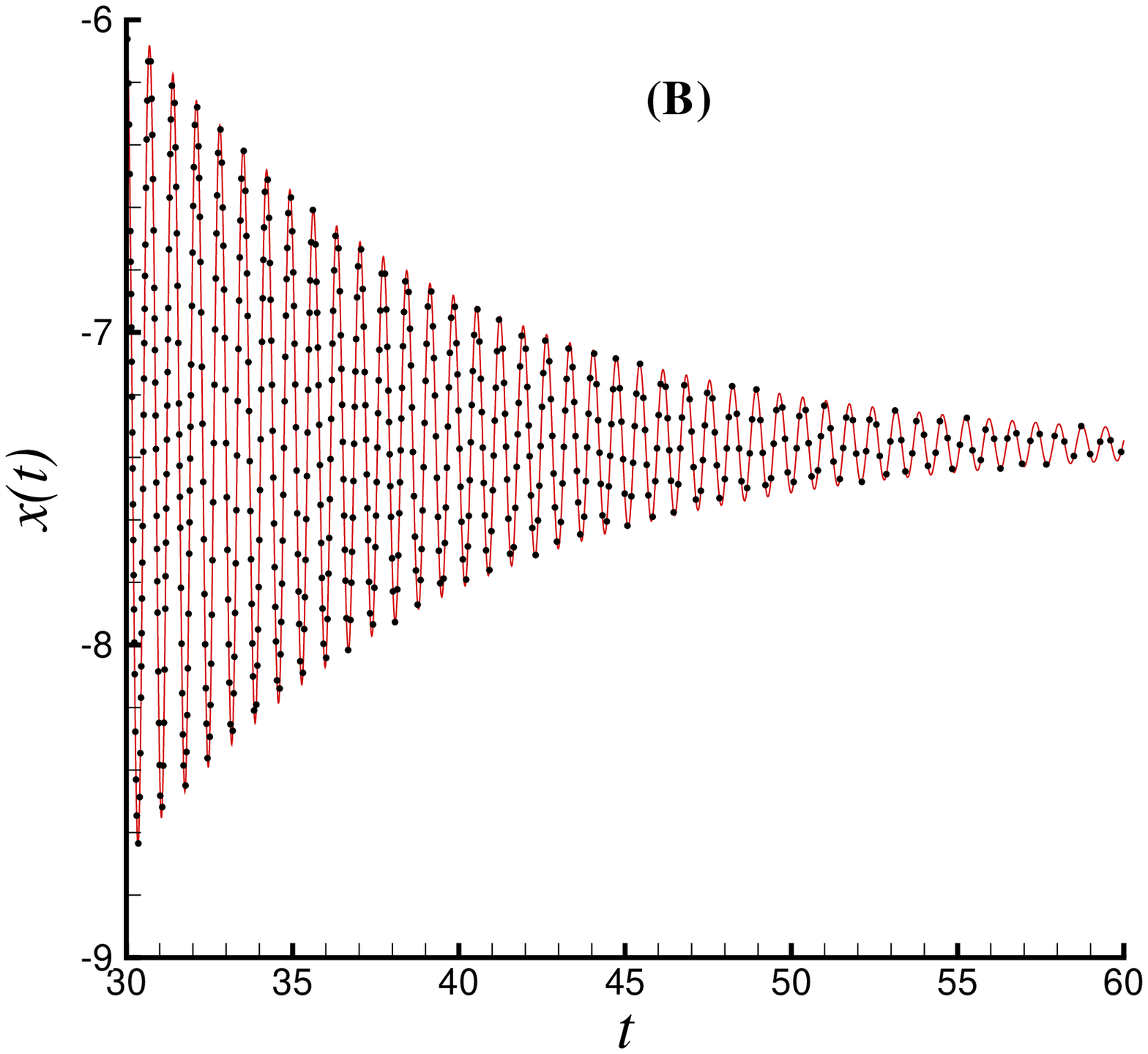}

\caption{(A) Comparison of $x(t)$ in case of $\sigma=10, R = 21.5,
b =-8/3$ with the initial state $x=y=z=5$ by means of $\tau=0.01$
LTU and the $M$th-order scheme (\ref{Taylor-procedure}) based on
truncated Taylor series in the interval $0 \leq t \leq 30$. Line:
$M=30$; Symbols: $M=40$. (B): the same, but in the interval $30
\leq t \leq 60$.} \label{figure:R21d5M30M40}
\end{figure}

Similarly, the so-called computational chaos (CC) and
computational periodicity (CP) mentioned by Lorenz
\cite{Lorenz1989,Lorenz2006} can be turned away, too.  For
example, when $\sigma=10, b =-8/3$ and $R = 21.5 < R_c$ with the
initial state $x = y =z =5$, it is found that the computed result
$x(t)$ given by the $M$th-order scheme (\ref{Taylor-procedure})
based on the truncated Taylor series with $\tau=0.01$ LTU keeps
chaotic when $M=2$ but tends to a positive fixed-point when $M=3$,
as shown in Figure \ref{figure:R21d5M2M3}.  Since $R = 21.5 <
R_c$, the chaotic solution given by $M=2$ is obviously wrong, and
the exact solution of $x(t)$ must tend to a fixed point for a
large enough time.  However, it is  {\em not} guaranteed whether
or not the exact solution of  $x(t)$ indeed tends to the positive
fixed point.  To get a reliable conclusion, computed results with
large enough critical predictable time $T_c$ are needed.  It is
found that the computed result given by $M=30$ agrees well with
that given by $M=40$ in the interval $0 \leq t \leq 100$, as shown
in Figure \ref{figure:R21d5M30M40}.  These two results, which are
reliable in $0 \leq t \leq 100$, clearly indicate that,  the
solution in case of $\sigma=10, b =-8/3$ and $R = 21.5 < R_c$ with
the initial state $x = y =z =5$ is {\em not} chaotic, and besides
$x(t)$ tends to a {\em negative} fixed-point. Thus, both of the
two results given by $M=2$ and $M=3$ are wrong: one gives the
so-called computational chaos from the result based on $M=2$, and
the other a wrong prediction from the result based on $M=3$.  In
this way, both of computational chaos and computational
uncertainty of prediction can be avoided by using reliable results
with a large enough critical predictable time $T_c$. Similarly,
the so-called computational periodicity (CP) can be avoided.

The above examples illustrate the importance and need of
introducing the concept of the critical predictable time $T_c$. In
this way, computational chaos (CC), computational periodicity
(CP), and computational uncertainty of prediction (CUP) of
nonlinear dynamic systems can be avoided by using reliable results
with a large enough critical predictable time $T_c$.

Mathematically, given a critical predictable time $T_c$, one can
always determine the order $M=T_c/3$ of the truncated Taylor
series (\ref{Taylor-procedure}), and the number $K=0.4 \; T_c$ of
the accurate decimal places of data by means of (\ref{tmax-M}) and
(\ref{tmax-K}) in case of $\sigma=10, R = 28, b =-8/3$ with the
initial condition $x(0)=-15.8, y(0)=-17.48, z(0)=35.64$, although
the needed CPU times might be rather long.  So, from pure
review-points of mathematics, using the concept of the critical
predictable time $T_c$, the computational uncertainty of
prediction of chaotic dynamical systems can be avoided, as long as
we have fast enough computer with large enough memory (RAM).

\section{On the origin of prediction uncertainty of chaos}

Unfortunately, most nonlinear dynamical systems describe physical
phenomena in nature.   Thus, results given by these models should
have physical meanings.   So, it is necessary to investigate the
prediction uncertainty of chaos from physical view-points.

The famous Lorenz equation (\ref{geq:lorenz:original}) is a
macroscopical model for climate prediction on earth: it  models a
unsteady flow occurring in a layer of fluid of uniform depth $H$
with a constant temperature difference $\Delta T$ between the
upper and lower surfaces, and $x(t)$ is proportional to the
intensity of convective motion \cite{Lorenz1963}.   So, it is
reasonable that the influence of physical factors in the level of
atom and molecule on the climate is neglected completely in Lorenz
equation.  On the one hand, for purpose of climate prediction,
measured data are  {\em unnecessary} to be precise in the level of
atom and molecule. On the other hand, as output of a macroscopical
model, computational results given by Lorenz equation are {\em
impossible} to be precise in the microcosmis level.

As mentioned in \S 3,  a reliable computational chaotic result
with $T_c=1200$ LTU is obtained by means of $M=400$ and $K=800$
with the  {\em exact} initial condition $x(0) = -15.8$, $y(0) =
-17.48$, $z(0) = 35.64$.  Note that $K = 800$ corresponds to very
precise data.   However, according to (\ref{tmax-M}) and
(\ref{tmax-K}), it is unnecessary to use such precise data to get
a chaotic result with $T_c=1200$ LTU.  Substituting $T_c = 1200$
into (\ref{tmax-M}) and (\ref{tmax-K}) give $M=400$ and $K = 480$,
respectively.  Thus, mathematically speaking, the precise data
with 480 decimal places {\em must} be used in order to get a
chaotic result reliable in the interval $0 \leq t \leq 1200$
(LTU).  This conclusion is obtained with the assumption that the
initial condition is exact.  Unfortunately, the initial condition
is not perfect in practice.  It is interesting that, substituting
$T_c = 1200$ into (\ref{tmax-Dx0}) gives $\Delta x_0 = 10^{-480}$.
This indicates that  the initial condition must be at least with
the same accuracy as all computed data used at different time.
Therefore, from pure view-points of mathematics, the initial
condition (and all computed data)  must be with the precision
$\Delta x_0 = 10^{-480}$ so as to get a reliable chaotic solution
with $T_c = 1200$ LTU.

First, to show how small the number $10^{-480}$ is, let us
 compare it with some physical constants.  According to
NASA's Wilkinson Microwave Anisotropy Probe (WMAP) project,  the
age of the
universe\footnote{http://en.wikipedia.org/wiki/Universe} is
estimated to be about $ 1.373  \times 10^{10}$ years, i.e. $T_u
\approx 4.33 \times 10^{17}$ seconds, and its diameter is about 93
billion light years, i.e. $d_u \approx 8.8 \times 10^{26}$ meter.
On the other side, helium is the smallest atom with a radius of 32
picometer\footnote{http://en.wikipedia.org/wiki/Atom\#Size}, i.e.
$r_a \approx 3.2 \times 10^{-11}$ meter, and the diameter of the
nucleus for a proton in light hydrogen is about 1.6
feotometer\footnote{http://en.wikipedia.org/wiki/Atomic\_nucleus},
i.e. $d_n \approx 1.6 \times 10^{-15}$ meter.  Assume that one
object ``moves'' a distance of radius of helium or a diameter of a
proton in light hydrogen since the beginning of the universe, i.e.
the Big Bang\footnote{http://en.wikipedia.org/wiki/Big\_Bang}.
Then, the corresponding velocities are $u_a = r_a/T_u \approx 7.39
\times 10^{-29}$ (meter/second) and $u_n = d_n/T_u \approx 3.7
\times 10^{-33}$ (meter/second), respectively.  However, even
dividing them by the speed of
light\footnote{http://en.wikipedia.org/wiki/Light\_speed} $c
\approx 3.0 \times 10^9$ (meter/second) that is assumed to be the
largest velocity in nature, we have only the dimensionless
velocities $\bar{u}_a = r_a/(c T_u) \approx 2.46 \times 10^{-38}$
and $\bar{u}_n = d_n/(c T_u) \approx 1.23 \times 10^{-42}$,
respectively.  Even so, they are {\em much} larger than
$10^{-480}$, because both $\bar{u}_a \div 10^{-480} = 2.46 \times
10^{442} $ and $\bar{u}_n \div 10^{-480} = 1.23 \times 10^{438}$
are {\em much} greater  even than $d_u / d_n = 5.5 \times
10^{41}$, the ratio of the diameter of the universe to the
diameter of the nucleus for a proton in light hydrogen!

Secondly, according to the Heisenberg uncertainty principle in
quantum physics \cite{Heisenberg1927}, the values of certain pairs
of conjugate variables (position and momentum, for instance)
cannot both be known with arbitrary precision, and any measurement
of the position with accuracy $\Delta \delta$ and the
 momentum with accuracy $\Delta p$ must satisfy
 \begin{equation}
 \Delta \delta \; \Delta p  \geq  \frac{h}{4\pi}, \label{UncertaintyPrinciple}
 \end{equation}
 where $h = 6.62606896 \times 10^{-34}$ [J] [S] is Planck's
 constant\footnote{http://en.wikipedia.org/wiki/Planck\_constant}
 with the unit [J] of energy (Joule) and the unit [S] of time
 (Second).  Rewriting $\Delta p = m \Delta v$, where $m$ denotes the
 mass and $v$ the velocity, one has
 \begin{equation}
\Delta v  \Delta \delta \geq \frac{h}{4\pi m }.
 \end{equation}
Therefore, the more precisely the velocity $v$ is known, the less
precisely the position $\delta$ is known.  Because Lorenz equation
models the flow of fluid on the earth, the worst measurement of a
position is with accuracy $\Delta \delta = d_E$, where $d_E =
1.2745 \times 10^7$ (meter) is the average diameter of the earth.
Then, the most precise measurement of velocity is at most
\begin{equation}
\Delta v  \geq \frac{h}{(4\pi d_E) m} = \frac{4.1372 \times
10^{-42}}{m}.
\end{equation}
 Even if $m$ is regarded as the mass of earth,
i.e. $m = 5.9742 \times 10^{24}$ (kg), the most precise
measurement of velocity is at most
\begin{equation}
\Delta v \geq 6.92511 \times 10^{-67} (m/s).
\end{equation}
Let  $\bar{v}$ denote the dimensionless velocity and $U$ the
velocity reference, respectively.  The above formula gives
\[   \Delta \bar{v} \geq  \frac{6.92511 \times 10^{-67}}{U}.   \]
According to the general relativity, light propagates fastest in
nature.  However, even if the velocity of light is used as the
reference velocity, i.e. $U \approx 3.0 \times 10^9$ (m/s), the
most precise measurement of dimensionless velocity on earth is at
most
\begin{equation}
\Delta \bar{v} \geq  2.3 \times 10^{-76}.
\end{equation}
Therefore, it is impossible to measure a dimensionless velocity
more precise than the above value.  Here, it should be emphasized
that the above very tiny number $2.3\times 10^{-76}$ is even {\bf
much} larger than $10^{-480}$: the ratio $ (2.3 \times 10^{-76})
\div 10^{-480} = 2.3 \times 10^{108}$ is much
 greater even than $d_u / d_n = 5.5 \times
10^{41}$, the ratio of the diameter of the universe to the
diameter of the nucleus for a proton in light hydrogen!
Therefore, according to the Heisenberg uncertainty principle in
quantum physics,  it is {\em physically} impossible to give an
initial condition with the precision $\Delta x_0 = 10^{-480}$,
which is however {\em mathematically} necessary to get a chaotic
result reliable in the interval $0 \leq t \leq 1200$ (LTU), as
mentioned in \S 3.

How can we interpret the above interesting result?  It seems
unavoidable to use nonlinear dynamical models to describe the
nature, and besides chaos generally exist in various nonlinear
dynamical models. However, as mentioned above, in order to get
chaotic results reliable in a long enough time, we need initial
condition with precision even higher than the most precise
measurement allowed by the Heisenberg uncertainty principle in
quantum physics.   Note that the precision, which is {\em
mathematically} necessary for the initial condition and all
computed data at different time, is so high that even the
quantum-fluctuation  becomes a very important physical factor and
therefore can not be neglected.  Therefore, the famous
``butter-flyer effect'' of Lorenz equation should be replaced by
the so-called ``quantum-fluctuation effect'': even the {\em
microcosmic} physical uncertainty such as quantum-fluctuation may
produce a large variations in the long-term macroscopical behavior
of a chaotic dynamic system describing natural phenomena.

Thus, although from mathematical points of view we can indeed
obtain reliable chaotic solution with $T_c
> 1200$ LTU in case of $\sigma=10, R = 28, b =-8/3$ with the exact initial
condition $x(0)=-15.8, y(0)=-17.48, z(0)=35.64$ by means of $\tau
= 0.01$, $M=400$ and $K = 480$, unfortunately, this {\em
mathematical} solution with such a high precision  has no {\em
physical} meanings. It should be emphasized that Lorenz equation
is a {\em macroscopical} model for climate prediction, and thus
{\em microcosmic} physical factors such as the quantum-fluctuation
are neglected. However, it is {\em mathematically} necessary for
Lorenz equation to have initial condition with such a high
precision that the Heisenberg uncertainty principle in quantum
physics must be considered {\em physically}.  This provides us a
``precision paradox of chaos''.

A paradox often brings us much deeper understandings about some
 thoughts and/or theories.  What can such a paradox tell
us? It seems that, to avoid this paradox, the following
assumptions or view-points should be accepted:
\begin{enumerate}
    \item[(A)] Chaos is physically unpredictable.  The origin of the unpredictability of chaos comes
essentially from the microcosmic uncertainty, which is described
by the Heisenberg uncertainty principle in quantum physics;
    \item[(B)] Even macroscopical phenomena might be essentially uncertain,
and thus it would be more reliable and more economic to describe
them by probability;
    \item[(C)] Most of nonlinear dynamical models, which describe complicated macroscopical
    phenomena such as chaos and turbulence,  do not consider the influence of microcosmic uncertainty, and thus should
be modified.
\end{enumerate}

\begin{figure}
\centering\setcaptionwidth{5.0in}
\includegraphics[width=3.5in,angle=0]{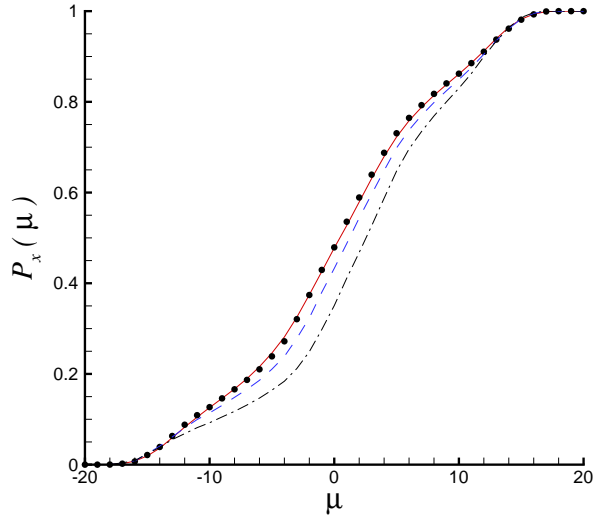}
\caption{The probability $P_x(\mu)$ in case of $\sigma=10, R = 28,
b =-8/3$ and the exact initial condition $x(0)=-15.8, y(0)=-17.48,
z(0)=35.64$ by means of $\tau = 0.01$ (LTU) with different $T_c$.
Symbols: result given by reliable solution with $T_c = 300$ (LTU);
Solid line: result given by reliable solution with $T_c = 75$
(LTU); Dashed line: result given by reliable solution with $T_c =
50$ (LTU); Dash-dotted line: result given by reliable solution
with $T_c = 25$ (LTU).} \label{figure:Px}
\end{figure}

To support the above interprets for the so-called
``precision-paradox of chaos'', let us consider the statistic
probability of $x(t)$ less than $\mu$, denoted by $P_x(\mu)$. The
probabilities $P_x(\mu)$ obtained by reliable chaotic results with
different critical predictable time $T_c$ in case of $\sigma=10, R
= 28, b =-8/3$ with the exact initial condition $x(0)=-15.8,
y(0)=-17.48, z(0)=35.64$ by means of $\tau = 0.01$ (LTU) are as
shown in Figure \ref{figure:Px}. Note that the probability
$P_x(\mu)$ given by $T_c=75$ LTU agrees well with the probability
given by the reliable computational result with $T_c = 300$ LTU.
It indicates that one can obtain a ``stable'' or ``convergent''
probability $P_x(\mu)$ by means of a reliable result with a proper
$T_c$ that is unnecessary to be very long. Note that it is much
easier to get a reliable chaotic result with $T_c=75$ LTU than
that with $T_c=1200$ LTU!  Therefore, it is much cheaper to get a
reliable probability $P_x(\mu)$ than a reliable time-series $x(t)$
with $T_c = 1200$ LTU.  So, it seems more reliable and especially
more ``economic'' to describe chaotic phenomena by means of
probability.   This partly supports our above-mentioned interprets
about the so-called ``precision-paradox of chaos''.

Thus, although the computational uncertainty of chaos can be
avoided from the mathematical points of view, it is unavoidable
from the physical points of view:  the so-called
``precision-paradox of chaos'' suggests us that the origin of the
uncertainty of chaos comes from the Heisenberg uncertainty
principle in quantum physics and thus is not avoidable, forever!

\section{Discussions}

In this paper a new concept, namely the {\em critical predictable
time} $T_c$, is introduced to give a more precise description of
computed chaotic solutions of nonlinear differential equations:
computed chaotic solutions are regarded to be reliable only when $
0 < t \leq  T_c$. This provides us a method or strategy to detect
the reliable result from a given computed solution. Besides, it
provides us a time scale for the so-called ``long-term'': $t$ is
regarded to be long-term as long as $t > T_c$.   It is also
suggested that numerical results beyond the critical predictable
time $T_c$ are unpredictable, and thus all related conclusions
based on computed chaotic results beyond the critical predictable
time $T_c$ are unreliable and doubtable. In this way, the
numerical phenomena such as computational chaos (CC),
computational periodicity (CP) and computational prediction
uncertainty, which are mainly based on long-term properties of
computed results,  can be avoided, as shown in \S 4. By means of
this concept, the famous conclusion ``accurate long-term
prediction of chaos is impossible'' should be replaced by a more
precise conclusion that ``accurate prediction of chaos beyond the
critical predictable time $T_c$ is impossible''.

For a nonlinear dynamic system with chaotic behavior, one had to
solve it by at least two different computation schemes so as to
get the critical predictable time $T_c$.  Certainly, it is better
to use more different computation schemes  to investigate the
reliability of computed results with chaos: the more, the better.
So, the reliability of chaotic solutions is a relative concept: it
is dependent on not only  nonlinear differential equations but
also the accuracy of initial guess, time-step and computation
schemes.  Without knowing the exact solution, such kind of
reliable solutions within the critical predictable time $T_c$
might be the best in practice: they are at least predictable, i.e.
different computation schemes lead to very close results.  Note
that even the definition (\ref{maximum-time-criteria}) of the
critical predictable time is dependent upon the two constants
$\delta$ and $\epsilon$. Fortunately, the same qualitative
conclusions can be obtained even by different (but reasonable)
values of $\delta$ and $\epsilon$. So, all of our conclusions
mentioned in this paper have general meanings.

On the one hand, the so-called critical predictable time $T_c$
provides us a scale to investigate chaos more precisely. On the
other hand, the symbolic computation software (such as
Mathematica) provide us a convenient way to investigate the
influence of truncation-error, round-off error, and inaccuracy of
initial guess on the critical predictable time $T_c$.   It is
found that $T_c$ is directly proportional to $M$, the order of the
truncated Taylor series (\ref{Taylor-procedure}), and $K$, the
number of  decimal places of all computed data. Besides, the
precision of initial conditions must increase exponentially as
$T_c$ enlarges. For example, in case of $\sigma=10, R = 28, b
=-8/3$ with the initial condition $x(0)=-15.8, y(0)=-17.48,
z(0)=35.64$, we obtain a reliable chaotic result with $T_c = 1200$
LTU by means of $\tau = 0.01$ (LTU), $M = 400$ and $K=800$.   Such
a reliable chaotic solution in so long time is reported for the
first time. Mathematically, given a critical predictable time
$T_c$, we can always get a reliable chaotic result in the interval
$0 \leq t \leq T_c$ by means of a high-performance computer with
large enough memory (RAM) and fast enough CPU, although the needed
CPU time might be rather long. Therefore, in essence, the
computational uncertainty of prediction for chaos can be avoided,
if only from the mathematical points of view.

However, the precision of initial condition and computed data at
each time-step needed for a large $T_c$ (such as $T_c=1200$ LTU)
is {\em mathematically}  so high that such precise data is {\em
physical} impossible due to Heisenberg uncertainty principle in
quantum physics.  Note that the precision, which is {\em
mathematically} necessary for the initial condition and all
computed data at different time, is so high that even the
quantum-fluctuation  becomes a very important physical factor and
therefore can not be neglected.  But, as a macroscopical model for
climate prediction on earth, Lorenz equation  completely neglects
the influence of physical factors in the level of atom and
molecule on the climate.  This provides us the so-called
``precision-paradox of chaos'', which implies that the prediction
uncertainty of chaos is physically unavoidable and besides even
the macroscopical phenomena might be essentially stochastic and
thus should be described  by probability more economically.

Many nonlinear evaluation equations for macroscopical phenomena,
such as Lorenz equation for climate prediction and Navier-Stokes
equation for turbulent viscous flows, completely neglect the
influence of physical factors in the level of atom and molecule.
However, the so-called ``precision paradox of chaos'' suggests
that this might be wrong: the influence of physical factors in the
level of atom and molecule should be considered for complicated
nonlinear dynamic systems.  It is well-known that turbulent flows
are much more complicated than chaos.  So, one should be very
careful to apply numerical schemes to investigate turbulent flows.
Currently, the DNS (direct numerical simulation) is frequently
used in computational fluid dynamics (CFD) to simulate turbulence
flows
\cite{JFM1997-Le,AnnReview1998-Moin,AnnReview1999-Scardovelli,PhysicsFluids1999-Moser,JCP-2006Martin}.
However, it is a pity that the mathematical sensitivity of DNS
results for turbulent flows to the inaccuracy of initial
conditions, round-off error and truncation error has not been
studied systematically, mainly because the DNS is rather
time-consuming. Without a method or strategy to detect the
reliability of a given DNS result, we has many reasons to assume
that something without physical meanings (similar to computational
periodicity and  computational chaos mentioned by Lorenz
\cite{Lorenz1989, Lorenz2006}) might be contained in the so-called
DNS ``solutions'' for turbulence, and thus conclusions based on
such kind of unreliable computed results might be doubtable.  More
importantly, all models for turbulent flows completely neglect the
influence of microcosmic factors in physics. And this might be one
of the reasons why there is no a satisfactory model  to describe
 all turbulent flows precisely.

Note that the concept of the critical predictable time $T_c$ is
not new: it is rather similar to the so-called ``critical
decoupling time'' mentioned by Teixeira {\em et al}
\cite{Teixeira2007}.  However, this concept has never been
 obtained enough recognition.  In this paper, we shaw the importance
of such kind of concept for the reliability of computed chaotic
results, and also for the avoidance of computational prediction
uncertainty, computational chaos (CC) and computational
periodicity (CP).  More importantly, the concept of critical
predictable time $T_c$ greatly deepens and enriches our
understanding about chaos, not only mathematically but also
physically.

Nonlinear dynamical systems describing chaos or turbulence might
be much more complicated than we thought: we should feel awe to
them. It is the time for us to consider seriously the reliability
of a mass of computed chaotic or turbulent results reported every
day!

\hspace{-0.5cm} {\bf Acknowledgement} The author would like to
express his sincere thanks to Dr. L.S. Yao, Dr. D. Hughes, and Dr.
J. Teixeira for providing some valuable references.  Thanks to the
anonymous referees for their valuable comments.

\bibliographystyle{unsrt}

\end{document}